\documentclass[aps,12pt,axodraw,nofootinbib,superscriptaddress,aps]{revtex4}
\pdfoutput=1
\usepackage{epsfig}
\usepackage{amsmath}
\usepackage{bm}
\usepackage{times}
\usepackage{graphicx}
\usepackage{epstopdf}
\usepackage{amsfonts}
\usepackage{bm}
\usepackage{epsfig}
\usepackage{graphics}
\usepackage{xspace}
\usepackage[usenames]{color}

\def\C{\mathcal{C}}
\def\bma#1{\mbox{\boldmath{$#1$}}}
\def\nn{\nonumber}
\def\bea{\begin{eqnarray}}
\def\eea{\end{eqnarray}}
\def\ba{\begin{eqnarray}}
\def\ea{\end{eqnarray}}
\def\be{\begin{equation}}
\def\ee{\end{equation}}

\def\beq{\begin{equation}}
\def\eeq{\end{equation}}

\def\nn{\nonumber}

\begin{document}

\title{\Large Probing for  Invisible Higgs Decays with Global Fits}
\author{J.R. Espinosa}
\affiliation{ICREA at IFAE, Universitat Aut{\`o}noma de Barcelona, 08193 Bellaterra, Barcelona, Spain}

\author{C. Grojean}
\affiliation{Theory Division, Physics Department, CERN, CH-1211 Geneva 23, Switzerland}

\author{M. M\"uhlleitner}
\affiliation{Institute for Theoretical Physics, Karlsruhe Institute of Technology, D-76128 Karlsruhe, Germany}

\author{M. Trott}
\affiliation{Theory Division, Physics Department, CERN, CH-1211 Geneva 23, Switzerland}

\date{\today}
\begin{abstract}
We demonstrate by performing a global fit on Higgs signal strength data that large invisible branching ratios (${\rm Br}_{inv}$) for a Standard Model (SM) Higgs particle are currently consistent with the experimental
hints of a scalar resonance with mass $m_h \sim 124 \, {\rm GeV}$. 
For this mass, we find ${\rm Br}_{inv} < 0.64$ ($95 \%$ CL) from a global fit to individual channel signal strengths supplied by ATLAS, CMS and the Tevatron collaborations. Novel tests that can be used to improve the prospects of experimentally discovering the existence of a ${\rm Br}_{inv}$ with future data are proposed. These tests are based on the combination of all visible channel Higgs signal strengths, and allow us to examine the required reduction in experimental and theoretical errors in this data that would allow a more significantly bounded invisible branching ratio to be experimentally supported. We examine in some detail how our conclusions and method are affected when a scalar resonance at this mass scale has couplings deviating from the SM ones.
\end{abstract}
\maketitle
\section{Introduction}
Two outstanding questions of importance that the LHC should shed light on are the origin of
electroweak symmetry breaking (EWSB), and the relationship 
of the mechanism of EWSB to new states beyond the Standard Model (SM).

There is strong indirect evidence for the EWSB sector being described by a theory that includes a particle that (at least) approximately has the properties of the SM Higgs boson. This evidence follows from many observables in flavour physics, from electroweak 
precision data (EWPD), LEP, the Tevatron and now the LHC. The SM Higgs is consistent with the results of these experimental probes
in its pattern of breaking custodial symmetry ($\rm SU(2)_c$)  \cite{Susskind:1978ms,Weinberg:1979bn,Sikivie:1980hm}
as well as in the manner by which it sources the experimentally established pattern of flavour violation. 
In light of these results, there is substantial indirect evidence that a scalar field involved in EWSB will also be SM Higgs like in that   
the soft Higgs theorems of Refs.~\cite{Shifman:1979eb,Vainshtein:1980ea}  will be approximately respected, i.e. the scalar field
will couple to the SM fields with a strength that is proportional to the mass of the corresponding SM particle.

Directly, LEP, the Tevatron and LHC have jointly excluded large regions of possible Higgs masses in the SM. The upper bound (in the low mass region)
for the SM Higgs is now restricted to $m_h <130 \, {\rm GeV}$ by ATLAS \cite{atlasnote2012019} and $m_h <129 \, {\rm GeV}$ by CMS \cite{CMSnote12008} at $99 \, \%$ CL, with a suggestive clustering of possible signal events around $m_h \sim 124 \, {\rm GeV}$.
However, in spite of these results, the Higgs hypothesis is not yet established. In particular, there remains a significant freedom in the
allowed couplings of a scalar effective field to the SM gauge bosons and fermions - so long as such a resonance has the approximate symmetries and properties discussed above \cite{Carmi:2012yp,Azatov:2012bz,Espinosa:2012ir}.\footnote{For other model independent approaches to Higgs couplings determination see \cite{Lafaye:2009vr,Englert:2011aa,Klute:2012pu}.} 
Such deviations in the properties of a scalar field from the SM Higgs can be interpreted as following from the Higgs boson emerging from a strongly interacting sector as a pseudo-Goldstone boson, or
as the leading effect in the effective theory of more massive states that are integrated out.

In light of this experimental situation, attempting to use current (and future) Higgs signal strength parameters to establish relationships between the EWSB sector and
beyond the SM states is speculative. This is certainly the current status, as the suggestive clustering of signal excesses at $m_h \sim 124 \, {\rm GeV}$
has not risen to the level of experimental evidence for a scalar resonance. Nevertheless, in this paper we will assume that
future data will support the discovery of a scalar resonance at approximately this mass scale. Further, we will consider current signal strength measurements as indicative of the
properties that such a scalar resonance has when performing global fits. In anticipation of such a discovery, it is of interest to
consider how to efficiently extract evidence of yet other states coupled to such a scalar field.

The gauge invariant mass operator of the scalar degrees of freedom, being of dimension two, is expected to couple
generally to all degrees of freedom. It is difficult to forbid a coupling of this operator at the renormalizable (or non-renormalizable) level to new states.
As such, the measurement of the decay width of a new scalar resonance to states that do not directly lead to significant excesses in the Higgs discovery channels, 
defined in this paper as its invisible branching ratio ${\rm Br}_{inv}$, could be the first direct measurement of interactions with states beyond the SM.
This exciting possibility has lead to many studies on extracting the invisible width of the Higgs. In this paper we will explore a very straightforward route using current and future experimental results on Higgs properties, expressed in terms of signal strength data,
to probe for evidence of a ${\rm Br}_{inv}$. We first show (in Section \ref{SMHiggs1}) that the current experimental hints of a new scalar resonance at $m_h \sim 124 \, {\rm GeV}$
do not put strong constraints on ${\rm Br}_{inv}$. We then explore in detail how to extract evidence for (or exclude) a ${\rm Br}_{inv}$ using global combinations 
of best fit signal strength parameters, performing global $\chi^2$ fits, and demonstrating a global probability density function (PDF) approach that can be used to explore and optimize searches for ${\rm Br}_{inv}$ 
in certain scenarios of beyond the SM (BSM) physics (Section \ref{PDFtest}).
In Section \ref{PDFtest2}, we examine the related issue of the precision with which ${\rm Br}_{inv}$ is expected to be known in these scenarios when errors are small enough for a resonance discovery to be claimed.

These promising results raise the question of the robustness of such a global approach. These techniques are most promising when BSM physics couples to the SM primarily
through the `Higgs portal' \cite{Schabinger:2005ei,Patt:2006fw,Chang:2008cw,MarchRussell:2008yu,Espinosa:2007qk,Pospelov:2011yp}, i.e. they are optimal in BSM scenarios where new states are not charged under the
SM gauge group and couple to the SM (initially) through the SM gauge singlet scalar mass operator. This is the case we explore in detail throughout Section \ref{SMHiggs1}.
We briefly discuss and summarize the prospects for global fits to uncover ${\rm Br}_{inv}$ in broader scenarios in Section \ref{global} where the effective couplings of the Higgs to the SM fields
deviate from their SM values due to the Higgs being a pseudo-Goldstone boson or due to the presence of higher dimensional operators.
We present detailed numerics for these scenarios (based on current
global Higgs fits to experimentally reported signal strengths) throughout the remainder of Section \ref{global}.  

We find that global fits to signal strength parameters will be a powerful approach to search for evidence of ${\rm Br}_{inv}$ in the scenarios we consider.
But we note that challenges will exist in disentangling other new physics effects.  This will likely require a combination of indirect global fit approaches to extracting ${\rm Br}_{inv}$, which is  the focus of this paper,
and more traditional direct searches for ${\rm Br}_{inv}$ based on kinematic properties of Higgs signal channels. We compare and contrast these approaches in Section \ref{direct} and then
discuss our overall conclusion in Section \ref{concl}.

\section{The Standard Model Higgs and ${\bma{\rm Br}}_{\bma{inv}}$}\label{SMHiggs1}
In this section, we will present the current global best fit\footnote{
See  Ref. \cite{Espinosa:2012ir} for a detailed discussion on our fitting procedure. The values of the fifteen input $\hat{\mu}_i$ used are reported in the Appendix for completeness. Note that throughout this paper we will assume 
a $3 \%$ contamination due to $gg$ events in the $\gamma \gamma j \, j$ signal strength, see the Appendix for further details.} results for ${\rm Br}_{inv} = \Gamma_{inv}/( \Gamma_{inv} + \Gamma_{\rm SM})$ for the SM Higgs,
where $\Gamma_{inv}$ is the decay width to `invisible' states, as defined above, and $ \Gamma_{\rm SM}$ is the decay width of the SM Higgs.
We perform a global fit to the available Higgs signal data, fitting to fifteen Higgs signal strength parameters $\mu_i$ reported by ATLAS, CMS and the Tevatron collaborations which are defined as
\bea
\mu_i = \frac{[\sigma_{j \rightarrow h} \times {\rm Br}(h \rightarrow i)]_{observed}}{[\sigma_{j \rightarrow h} \times {\rm Br}(h \rightarrow i)]_{SM}}\ ,
\eea 
for a production of a Higgs  that decays into the visible channel $i$. We use the best fit values of $\mu_i$, denoted by $\hat{\mu}_i$, as reported by the experimental collaborations.
The label $j$ in the cross section, $\sigma_{j \rightarrow h}$,  is to denote that signal events in some final states are defined (by selection cuts) to
only be summed over a subset of Higgs production processes $j$.
We construct a global $\chi^2$ measure on the $\hat\mu_i$ by defining the matrix $\C$ as the covariance matrix of the
observables, and $\Delta \, \theta_i=\mu_i-\hat\mu_i$ as a vector of the difference between the signal strength variable $\mu_i$ and the best fit value of the signal strengths\footnote{In a simple counting experiment the definition is $\hat{\mu}=(n_{obs}-n_b)/n_s^{SM}$, in terms of the observed number of events ($n_{obs}$), the number of background events ($n_b$) and the expected number of SM signal events ($n_s^{SM}$).},
\bea
\label{chi2}
\chi^2(\mu_i) = (\Delta \theta_m)^T  \, (\C^{-1})_{mn} \, (\Delta \theta_n)=\sum_{i=1}^{N_{ch}}\frac{(\mu_i-\hat\mu_i)^2}{\sigma_i^2}\; .
\eea
Here $i = 1 \cdots N_{ch}$, where $N_{ch}$ denotes the number of channels. 
The matrix $\C$ is taken to be diagonal with the square of
the $1 \, \sigma$ theory and experimental errors added in quadrature 
for each observable, giving the error $\sigma_i$ in the equation above. Correlation
coefficients (currently not supplied by the experimental collaborations) are
neglected. For the experimental errors we 
use $\pm$ symmetric $1 \, \sigma$ errors on the reported $\hat{\mu}_i$. For theory predictions of the $\sigma_{j \rightarrow h}$ and related errors, 
we use the numbers given on the webpage of the LHC
Higgs Cross Section Working Group \cite{Dittmaier:2011ti}. 
The minimum ($\chi^2_{min}$) is determined, and the $68.2 \% \, (1 \, \sigma), 95 \% \, (2 \, \sigma)$ best fit confidence level  (CL) regions are given by $\Delta \chi^2 < 1,
\,\,4$, respectively, for $\chi^2 = \chi^2_{min} + \Delta \chi^2$.
Here, the CL regions are defined by the cumulative distribution
function (CDF) for a one-parameter fit. 

\begin{figure}[tb]
\includegraphics[width=0.48\textwidth]{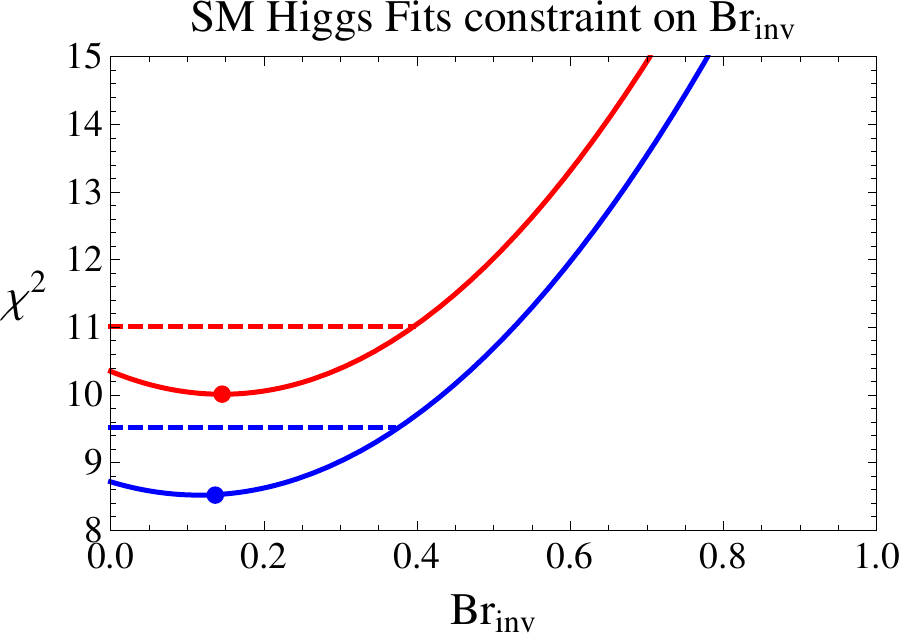}
\includegraphics[width=0.47\textwidth]{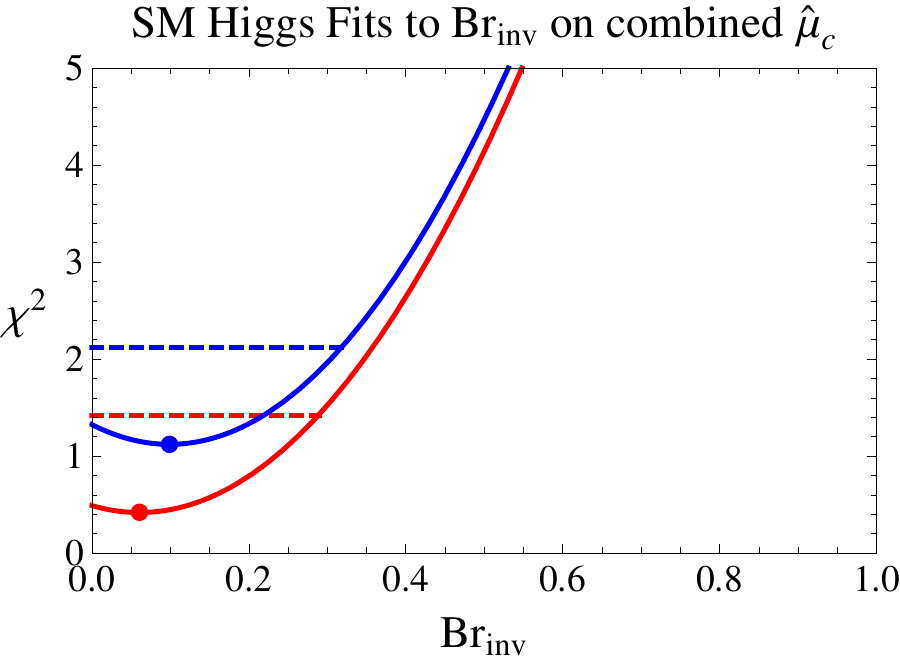}
\caption{\it Global fit to the best-fit signal strength parameters
in SM Higgs searches as supplied by the four experiments
for individual channels (left) and for their combinations (right). 
These results are based on post Moriond 2012 data (see the Appendix and Table I) when an invisible width is added to the SM as a free parameter. 
Left: The red (upper) solid curve is for $m_h = 125 \, {\rm GeV}$; the blue (lower) solid curve is for 
$m_h = 124 \, {\rm GeV}$. The one sigma region defined with the CDF for a one parameter fit is given by the horizontal dashed lines in each case and the best fit points are given by
$(m_h, {\rm Br}_{inv}) = (124,0.12),\, (125,0.15)$.
Right: The red (lower) solid curve is for $m_h = 125 \, {\rm GeV}$; the blue (upper) solid curve is for 
$m_h = 124 \, {\rm GeV}$. Now the best fit points are given by
$(m_h, {\rm Br}_{inv}) = (124,0.10),\, (125,0.06)$. Comparing these results gives a sense of the effect of neglected correlations amongst the individual signal channels in such fits.}\label{Fig.globalSM1}
\end{figure}

Although we are fitting for evidence of new `invisible' states,
we do not include effects due to new unknown interactions on the production of the SM Higgs in this section\footnote{In later sections, we examine the effect of contact interactions (that could be due to new states in a BSM sector)
on our conclusions and method. In this section, we are essentially assuming the case of new states that are not charged under the SM group, coupling primarily to the SM
through the SM gauge singlet operator $\rm H^\dagger \, H$.}.
We include an invisible width by modifying the SM branching ratios universally for each decay into final states $f$ via
\bea
{\rm Br}(h \rightarrow f) \equiv \frac{\Gamma(h \rightarrow f)}{\Gamma_{\rm
    SM} + \Gamma_{inv}} = (1 - {\rm Br}_{inv}) \times {\rm Br}_{SM}(h \rightarrow f).
\label{eq:brinv}
\eea
Thus, the effect of including an invisible width (of BSM origin) on the signal strengths is that the expected $\mu_i =1$ in the SM is modified to an expectation of $\mu_i = 1 - {\rm Br}_{inv}$. We fit for the parameter ${\rm Br}_{inv}$
assuming a SM Higgs with a total SM width $\Gamma_{\rm SM}$ and a particular Higgs mass. 
The resulting $\chi^2$ as a function of ${\rm Br}_{inv}$ is shown in Fig. \ref{Fig.globalSM1}.
Interestingly, we find that the global $\chi^2$ is minimized for a non-zero value of  ${\rm Br}_{inv}$.\footnote{Comparison of our results with those of Ref. \cite{Giardino:2012ww}, which finds
that a related global $\chi^2$ is minimized for ${\rm Br}_{inv} <0$, is not straightforward. For the results presented in this paper we use the $\hat{\mu}_i$ (and errors) as reported by the experimental collaborations. The results in Ref. \cite{Giardino:2012ww} are based on $\hat{\mu}_i$ constructed  from reported and expected CL limits that only approximate the experimentally reported $\hat{\mu}_i$, apparently introducing a distortion in the data (and associated errors) that affect the final conclusions. Of course, due to the large experimental errors at this time, the $95\%$ CL range is wide in our results and in Ref. \cite{Giardino:2012ww}.} An inspection of the data used for $m_h = 124 \, {\rm GeV}$ (see the Appendix) reveals $h\to \gamma\gamma$ (ATLAS) and $h\to WW$ (ATLAS and Tevatron) as the channels that most favor a nonzero
value of ${\rm Br}_{inv}$, while $h\to ZZ$ (ATLAS), $h\to \gamma\gamma$ (CMS) and $h\to b\bar{b}$ (Tevatron) are the channels which tend to drive ${\rm Br}_{inv}\rightarrow 0$.
This result also demonstrates that, despite a suggestive hint in the data for a Higgs like scalar resonance,  ${\rm Br}_{inv}$ remains essentially unconstrained for the SM Higgs in the current data set.
The allowed values are ${\rm Br}_{inv} < 0.37 (0.64)$ at $68 (95) \%$ CL respectively for $m_h = 124 \, {\rm GeV}$, and ${\rm Br}_{inv} < 0.39 (0.65)$ for $m_h = 125 \, {\rm GeV}$. 
If the ${\rm Br}^{min}_{inv} >0$ result (statistically marginal at this time) is confirmed by the future data set, and the Higgs is discovered,
future global fits of this form could be the first evidence of the Higgs coupling to new states.

It is instructive to look more closely at the $\chi^2$ fit to all channels we have performed rewriting 
Eq.~\ref{chi2} as
\be
\label{chi2n}
\chi^2(\mu)=\frac{(\mu-\hat\mu_c)^2}{\sigma_c^2} +\left[\sum_{i=1}^{N_{ch}}\frac{\hat\mu_i^2}{\sigma_i^2}-\frac{\hat\mu_c^2}{\sigma_c^2}\right]\ ,
\ee 
where we have introduced the combined variables 
\bea\label{csigma}
\frac{1}{\sigma_c^2}  = \sum_{i}^{N_{ch}} \frac{1}{\sigma^2_i}, \quad \quad  \frac{\hat{\mu}_c}{\sigma_c^2} = \sum_i^{N_{ch}} \frac{\hat \mu_i }{\sigma_i^2} \ . 
\eea
Note that Eq.~\ref{chi2n} is valid if all the $\mu_i$ are equal, as is the case for the SM with an addition of ${\rm Br}_{inv}$.
This decomposition illuminates what our analysis of the fit to individual channels really does.
The location of the minimum of the fit, and the $N \sigma$ intervals, is controlled by the first term
in Eq.~\ref{chi2n}, which depends only on the combined parameters $\hat\mu_c$ and $\sigma_c$
but not on the dispersion of the different $\hat\mu_i$'s around their average $\hat\mu_c$.
How good the fit is, is just given by the second piece in Eq.~\ref{chi2n}, which is simply $\chi^2_{min}$,
and does depend on how separate are the individual channel $\hat\mu_i$'s from $\hat\mu_c$.
Interpreting, as we do in this paper, deviations of $\hat\mu$ from its SM value of 1 as a Higgs invisible width,
one immediately obtains that the $\chi^2$ is minimized (defining ${\rm Br}^{min}_{inv}$) when
\bea
{\rm Br}_{inv}=1-\hat\mu_c.
\eea


This also offers the alternative approach of bypassing the individual channel analysis and using directly
the $\hat\mu_c$ values reported by the experiments in Table \ref{table:combineddata}. We can use this data to do the $\chi^2$ fit as the effect of  ${\rm Br}_{inv}$ on the signal strengths is a common multiplicative correction.
\begin{table}[h] 
\setlength{\tabcolsep}{5pt}
\center
\begin{tabular}{c|c|c|c|c} 
\hline \hline 
Experiments & $\hat\mu_c$, \, $m_h =124$ & $\sigma_c$,\, $m_h = 124$ & $\hat\mu_c$,\,  $m_h = 125$ & $\sigma_c$,\,  $m_h = 125$
\\
\hline
${\rm CMS}$ \, \cite{MPieri} & $0.98$ & $0.30$ & $0.94$ & $0.32$
\\
${\rm ATLAS}$ \, \cite{MPieri} & $0.61$ & $0.38$ & $0.81$ & $0.38$
\\
${\rm CDF \& D 0\! \! \! /}$  \, \cite{TEVNPH:2012ab}& $1.31$ & $0.60$ & $1.28$ & $0.62$
\\
\hline \hline
\end{tabular}
\caption{\it Combined signal strengths and errors from ATLAS, CMS and the Tevatron collaborations. Here we quote $\pm$ symmetric $1 \sigma$ errors.}
\label{table:combineddata} \vspace{-0.35cm}
\end{table}

These results lead to the combined values $\hat\mu_c=0.9,\sigma_c=0.22 \, (\hat\mu_c=0.94,\sigma_c=0.23) $ for $m_h = 124 \,(125) \, {\rm GeV}$.
This directly translates into the best fit results ${\rm Br}_{inv}^{\rm min}=0.09\pm 0.22 \, (0.06\pm 0.23)$ for $m_h=124 \, (125)\, {\rm GeV}$ (with the $95 \%$ CL limit ${\rm Br}_{inv} \leq 0.54 \, (0.52)$).
Comparing the best fit value with the results of our previous analysis in terms of the individual channels: $\hat{\mu}_c$ are $0.88 \, (0.85)$ for $m_h = 124, (125) \, {\rm GeV}$.
The results of the two fits, to the individual and to the combined $\hat\mu$'s are consistent within the quoted errors, indicating that neglected correlation effects in the individual
signal channel fits do not dramatically change the fit results we will show.
The slight preference for a nonzero invisible width is driven by ATLAS data at this time. 

\subsection{Global PDF approach to discovering  ${\bma{\rm Br}}_{\bma{inv}}$ in the SM}\label{PDFtest}

The $\chi^2$ approach we have discussed can be justified on the basis of a more detailed analysis that makes use of the combination of the PDF's for all sensitive Higgs search channels. To the extent that these PDF's are well described by Gaussian distributions, both approaches are basically equivalent in terms of the discovery reach afforded. The combined PDF approach however has a more direct physical interpretation and makes clearer the expected experimental sensitivity required to discover or bound ${\rm Br}_{inv}$. Moreover, this treatment is more powerful as it could also capture possible deviations from simple Gaussian shapes. On the other hand, the $\chi^2$ fit is useful to determine if a reduction in Higgs signal yields in all channels
is really universal and leads to a value of $\chi^2_{min}$ indicating a good fit, and is very convenient for taking into account the effect of imposing EWPD constraints.
In this section, we will discuss global searches from the global PDF perspective, discussing the relationship between the errors in $\hat{\mu}_i$ and the discovery reach (or exclusion prospects) for ${\rm Br}_{inv}$ with such an approach. 


As in the previous $\chi^2$ analysis, the key point to the power of global searches is the fact that,\footnote{In the absence of an indirect impact of new states on production or decay of the Higgs through induced operators.} 
in the presence of  ${\rm Br}_{inv}$, the expected measured values of  the strength of the signal with respect to pure SM expectations are modified as $(\mu_i = 1) \rightarrow (\mu_i = 1-{\rm Br}_{inv})$.  Due to this, one can construct
a global PDF (combining the PDF's of individual channels) sensitive to this shift in the signal strengths which aids in experimentally distinguishing ${\rm Br}_{inv} \neq 0$ from the SM case ${\rm Br}_{inv} = 0$ .
As in Ref.~\cite{Azatov:2012bz}, we will assume that the PDF for each $\mu_i$ reported by the experimental collaborations can be approximated by Gaussian distributions
\bea
pdf_i(\mu,\hat{\mu}_i,\sigma_i) \approx e^{-(\mu - \hat{\mu}_i)^2/(2 \sigma_i^2)},
\eea
with one sigma error $\sigma_i$, and best fit value $\hat{\mu}_i$ for the signal channels $i$. This is the case (especially for $\mu$ near $\hat\mu_i$) as long as the number of events is large, $ > O(10)$ events, \cite{Azatov:2012bz}  and systematic errors are subdominant. Using experimentally reported $\hat\mu_i$ and $\sigma_i$ is the best and simplest way of approximating the likelihoods in the neighbourhood of $\hat\mu_i$. Experimental information on the 95\% CL exclusion limits on $\mu_i$ give additional information on the PDF's up to higher values of $\mu$ and allow one to identify channels for which there are non-Gaussian tails. In such channels, using exclusion limits to extract the $\hat\mu_i$
(as done in Ref.~\cite{Giardino:2012ww}) will tend to overestimate
the $\hat\mu_i$. This would subsequently bias an extracted value of ${\rm Br}_{inv}$ based on such constructed $\hat\mu_i$.

A global combination of all the visible channels PDF's, where every PDF is approximated as above,
is obtained as a product of the $pdf_i(\mu)$ (where $i = 1 \cdots N_{ch}$) and it is also approximately Gaussian
\be
pdf(\mu,\hat{\mu}_c,\sigma_c) \propto  \prod_{i}^{N_{ch}} \, pdf_i(\mu,\hat{\mu}_i,\sigma_i)
= {\mathcal{N}_c} \, e^{-(\mu - \hat{\mu}_c)^2/(2 \, \sigma_c^2)}
\ee
where $\sigma_c$ and $\hat{\mu}_c$ are defined in Eq.~\ref{csigma} and
\bea
{\mathcal{N}_c} = \sqrt{\frac{2}{\pi \, \sigma_c^2}} \, \frac{1}{1 + {\rm Erf}(\hat{\mu}_c/\sqrt{2} \, \sigma_c)}.
\eea
Here we have normalized with the condition $\int_0^\infty \, pdf(\mu,\hat{\mu}_c,\sigma_c) d \mu = 1$ and ${\rm Erf}$ is the standard error function.
In the limit where $\sigma_i \approx \sigma$, and correlations are neglected, one has the simple approximation $\sigma_c \approx \sigma/\sqrt{N_{ch}}$.
By combining all of the $N_{ch}$ visible channels, the distinguishability of ${\rm Br}_{inv}>0$ from alternative hypotheses  (like pure background, or pure SM) is
improved due to this $\sim 1/\sqrt{N_{ch}}$ suppression of $\sigma_c$ compared to an individual visible signal channel's $\sigma_i$.
Assuming that with sufficient data the measured $\hat{\mu}_c$ converges to the theoretically expected $\mu_c = 1 - {\rm Br}_{inv}$, by constructing a combined PDF one can determine the value of $\sigma_c$ required 
to have the possibility to statistically pinpoint the presence of ${\rm Br}_{inv} \neq 0$.

\begin{figure}[tb]
\includegraphics[width=0.55\textwidth]{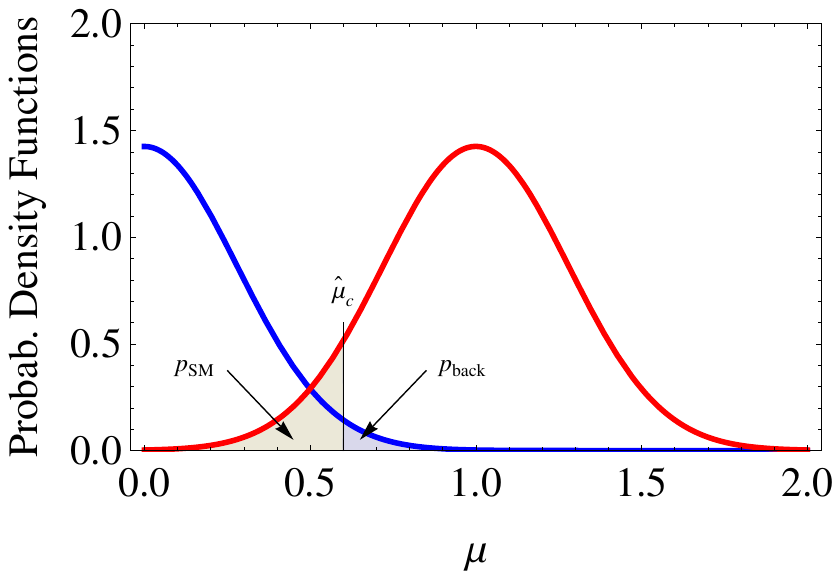}
\caption{\it Illustration of the probability density functions for the background-only (blue curve on the left) and SM (red curve on the right) and corresponding $p$-values for a hypothetical $\hat\mu_{c}=0.6$.}\label{Fig.pvalues}
\end{figure}
To find evidence of a nonzero ${\rm Br}_{inv}$ one has to be able to distinguish the ${\rm Br}_{inv}>0$ hypothesis (dubbed SMinv) from the SM Higgs hypothesis with ${\rm Br}_{inv} = 0$, and discern this case from the background-only hypothesis.
We use the same approach used routinely in experimental analyses to estimate the significance of a signal excess in the data, which quantifies how unlikely such an excess would be if interpreted as an upward fluctuation of the background. One defines a $p$-value for the background-only hypothesis as
\be\label{pback}
p_{back}=\int_{\hat\mu_c}^\infty pdf_b(\mu,\sigma_c)d\mu\ ,
\ee
where the background probability density function (or likelihood)
is approximately a Gaussian centered at $\mu=0$ (as $n_{obs}=n_b$) with some globally combined
1-standard deviation spread ($\sigma_c$) that results from the combination of $N_{ch}$ different channels each with an individual $\sigma_i$:\footnote{The overall normalization of $pdf_b(\mu,\sigma)$ can be fixed by $\int_0^{\infty} pdf_b(\mu,\sigma) = 1/2$, which follows from $\int_{-\infty}^{\infty} pdf_b(\mu,\sigma) = 1$. Note that negative $\mu$ in $pdf_b$ have a physical interpretation in terms of downward fluctuations of the background.}
\be\label{pdfback}
pdf_b(\mu,\sigma_c) \approx \prod_{i=1}^{N_{ch}} e^{-\mu^2/(2 \sigma_i^2)}\approx e^{-\mu^2/(2 \sigma_c^2)}\ .
\ee
For the $i^{th}$ channel, with an expected number of background events  $n_{b,i}$
and an expected number of signal events (in the SM) $n_{s,i}^{SM}$
one has  $\sigma_i=\sqrt{n_{b,i}}/n_{s,i}^{SM}$ (neglecting systematic effects).
Again, in the limit where all $\sigma_i$ are comparable, and correlations are neglected, one has the simple scaling $\sigma_c \approx \sigma_i/\sqrt{N_{ch}}$.
 A $p_{back}$-value as small as that corresponding to a 5$\sigma$ fluctuation will be required to claim Higgs discovery\footnote{The $p$-value corresponding to an $N \sigma$ fluctuation is $p_N=[1-{\mathrm{Erf}}(N/\sqrt{2})]/2$. One has $p_5=2.87\times 10^{-7}$.}. See Fig.~\ref{Fig.pvalues} for an illustration of the definition of the background $p$-value.

Besides having a small enough $p_{back}$ for a Higgs discovery, 
to claim evidence for ${\rm Br}_{inv} > 0$ we should be able to discard also the pure SM hypothesis (with ${\rm Br}_{inv}=0$), as a downward fluctuation in the signal yield could be misinterpreted as ${\rm Br}_{inv}>0$. We proceed exactly as before and construct the global PDF for the SM hypothesis as a Gaussian centred at
the SM value $\mu=1$:
\be\label{pdfSM}
pdf_{SM}(\mu,\sigma_c)\propto \prod_{i=1}^{N_{ch}} e^{-(\mu-1)^2/(2 \sigma_i^2)} = {\mathcal N}_{SM} \,  e^{-(\mu-1)^2/(2 \sigma_c^2)}\ .
\ee
Here ${\mathcal N}_{SM}$ is implicitly defined by the condition
\bea
\int_0^{\infty} pdf_{SM}(\mu,\sigma_c) = 1.
\eea
Also, neglecting systematic effects, $\sigma_i=\sqrt{n_{b,i}+n_{s,i}^{SM}}/n_{s,i}^{SM}$. 
 For $n_{s,i}^{SM}\ll n_{b,i}$ this is the same $\sigma_i$ as in Eq.~\ref{pdfback} so that we use the same notation for both.   We then compute the $p$-value associated with the pure SM hypothesis as
 \be\label{pSM}
p_{SM}=\int_0^{\hat\mu_{c}} pdf_{SM}(\mu,\sigma_c)d\mu\ .
\ee
See Fig.~\ref{Fig.pvalues} for an illustration. 

\begin{figure}[tb]
\includegraphics[width=0.45\textwidth]{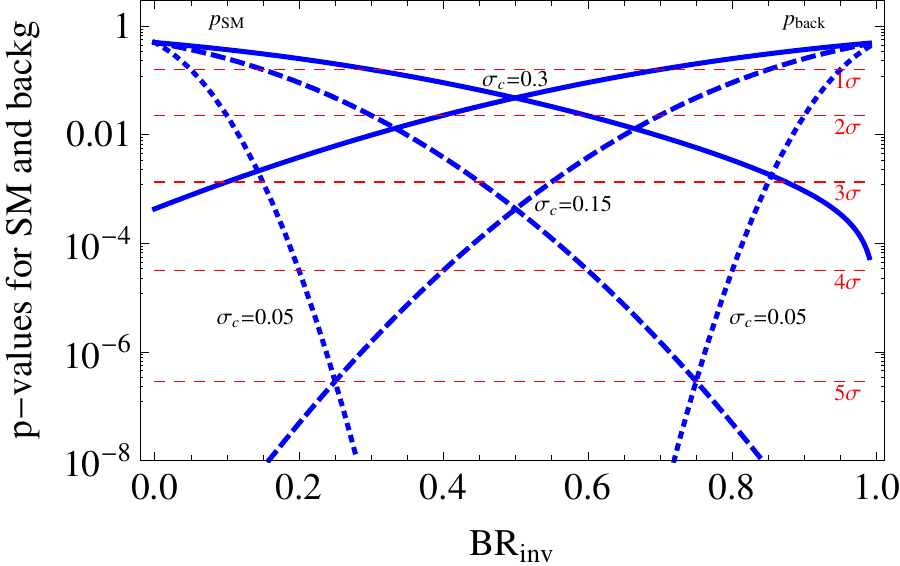}
\hspace*{0.5cm}
\includegraphics[width=0.45\textwidth]{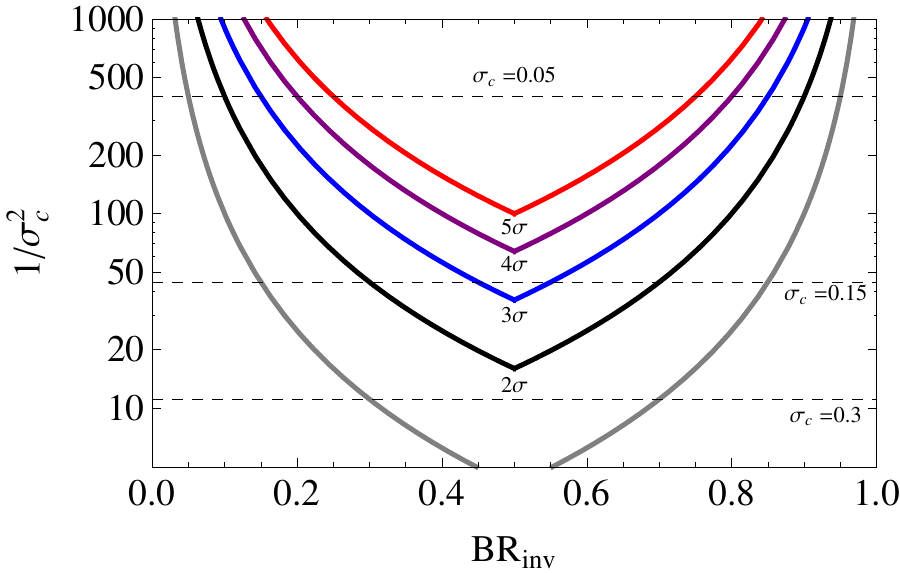}
\caption{\it Left: $p$-values for SM and background-only hypotheses (negative and positive-slope lines respectively) vs ${\rm Br}_{inv}$ for several values of the $1 \sigma$ error $\sigma_c$ on $\hat\mu$: its current value $\sigma_c=0.3$ (solid); half of it, $\sigma_c=0.15$, expected to be reached at the end of this year (dashed); and a future value $\sigma_c=0.05$ (dotted). The red dashed horizontal lines show the $p$-values corresponding to significances from $1 \, \sigma$ to $5 \sigma$'s.
Right: $\sigma_c$ (plotted as $1/\sigma_c^2$) required to pinpoint a non-zero ${\rm Br}_{inv}$ with a significance from 1 to 5 $\sigma$ (curves from lower gray to upper red). The precise condition imposed is that, for a given $\hat\mu_{c}=1- {\rm Br}_{inv}$, both $p_{back}$ and $p_{SM}$ (with $ {\rm Br}_{inv}=0$) are smaller than those corresponding to fluctuations from $1 \sigma$ to $5 \sigma$. The horizontal dashed lines show again the values $\sigma_c\simeq 0.3, 0.15$ and $0.05$. }\label{Fig.scBRinv}
\end{figure}

Claiming evidence for ${\rm Br}_{inv} > 0$ requires having simultaneously a small $p_{back}$ and $p_{SM}$. 
In order to quantify this, notice that $p_{back}\leq p_N$
requires $\hat\mu_c \geq N \, \sigma_c$ while $p_{SM} \leq p_N$ leads to 
$1-\hat\mu_c \geq  N \,  \sigma_c$. Using $\hat\mu_c =1-{\rm Br}_{inv}$,
we can obtain, as a function of ${\rm Br}_{inv}$, how small
$\sigma_c$ (the precision in the measurement
of $\hat\mu_c$) is required to be for a $N \sigma$ evidence of nonzero
Higgs invisible width. 
For reference, combining fifteen currently reported signal strengths from ATLAS, CMS, CDF and D0$\! \! \!/ \, \, $, while neglecting correlations, 
one finds $\sigma_c \simeq 0.3$ for $m_h = 124$ GeV. We estimate the current values of $\sigma_c$ per experiment as $\sigma_{c,{\rm ATLAS}}\simeq 0.5$, $\sigma_{c,{\rm CMS}}\simeq 0.4$ and $\sigma_{c,{\rm Tevatron}}\simeq 0.6$ by combining the individual channels reported in the Appendix experiment by experiment (while neglecting correlations) using Eq.~\ref{csigma}. This compares well to the combined $\hat{\mu}_c$ reported by the experimental collaborations
in Table \ref{table:combineddata}.
Figure~\ref{Fig.scBRinv}, left plot, shows
the $p$-values for both hypothesis for a ${\rm Br}_{inv}$ measurement with
certain precision $\sigma_c$, chosen for illustration at around its current 
value $\sim 0.3$ and future values $\sigma_c=0.15$ and $0.05$.
As expected, claiming evidence for ${\rm Br}_{inv}>0$ will be easier for ${\rm Br}_{inv}\sim 0.5$
and will be facilitated by a reduction in~$\sigma_c$. With the current value, the plot also shows that the weak indication of ${\rm Br}_{inv}\sim 0.12$ is perfectly compatible with a downward fluctuation of a 
SM-like Higgs, and even compatible with a background upward fluctuation at $\sim 3 \sigma$. Figure ~\ref{Fig.scBRinv} (right), shows the required precision $\sigma_c$ for $1 \sigma$ to $5 \sigma$ evidence of nonzero invisible width.  As expected, the ability to find evidence of a nonzero ${\rm Br}_{inv}$ degrades
for small values of this parameter, when it is harder to disentangle SMinv from the SM and also for ${\rm Br}_{inv} \rightarrow 1$, when it is hard to discern a small signal over background.\footnote{We have numerically cross checked the relationship between sensitivity to ${\rm BR}_{inv}$ and $\sigma_c$ shown in the p-value results with another 
simple test based directly on the lack of overlap of global PDF's. Introducing a PDF for the background only scenario, a SM PDF, and a test theory PDF, simply insisting that 
the $N$ sigma allowed $\mu$ in the test theory PDF lies outside of the $N$ sigma allowed regions of the other two PDF's, one finds a similar sensitivity to what is indicated for
a $2 \, N \, \sigma$ evidence for a common ${\rm BR}_{inv}$ in the $p$-value test.}
As $\sigma_i=\sqrt{n_{b,i}}/n_{s,i}^{SM}$, with more luminosity the statistical component of $\sigma_c$ will scale down with $ \sim 1/\sqrt{\cal{L}}$ (so that the plotted $1/\sigma_c^2$ increases linearly with $\cal{L}$). 
As an example, assuming this scaling of $\sigma_c$, and that the best fit value of ${\rm Br}_{inv} = 0.12$ obtained in the global fit is the true value, this indicates that accumulated signal events should be increased (compared to the current data set)
by a factor of $\approx 25 (100) $ to reach $2(4) \sigma$ evidence of this ${\rm Br}_{inv}$.

At the end of the current LHC run it is expected that the accumulated luminosity will be enough to reach the
level required for a $5 \sigma$ SM Higgs discovery per experiment.
This means that both $\sigma_{c,{\rm ATLAS}}$ and $\sigma_{c,{\rm CMS}}$ will
be down to $\sim 0.2$ or lower. (This expectation is consistent with recent public statements by CMS and ATLAS, see Ref.~\cite{MPieri}.) Taking such values for these quantities, and combining with the current $\sigma_{c,{\rm Tevatron}}$, we arrive at 
$\sigma_c\simeq 0.15$ (half the current value) as a reasonable 
number to expect by the end of the year.

\subsection{Bounding  ${\bma{\rm Br}}_{\bma{inv}}$ in the SM}\label{PDFtest2}

\begin{figure}[tb]
\includegraphics[width=0.55\textwidth]{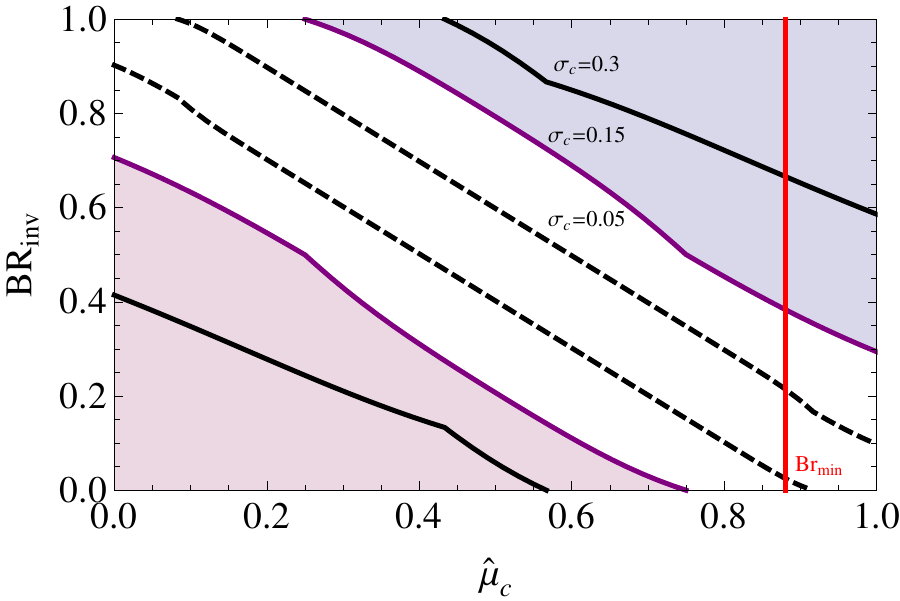}
\caption{\it 95\% CL exclusion limits for ${\rm Br}_{inv}$ as a function of the observed $\hat\mu_c$ for several values of its error $\sigma_c$: the current one ($\sigma_c=0.3$); the combined error that is estimated to be reached by both ATLAS and CMS at the end of the year ($\sigma_c=0.15$), with the 95\% CL excluded region shaded; and with a future $\sigma_c=0.05$. A red vertical solid line indicates the current  value of ${\rm Br}_{inv}$ obtained for $m_h = 124 \, {\rm GeV}$. The $95\%$ CL limit for $m_h = 124 \, {\rm GeV}$ obtained directly from the $\chi^2$ fit ($ < 0.64$) is consistent with the PDF test results shown.}\label{fig:BRinvLimits}
\end{figure}

With the measured overall $\hat\mu_{c}$, known with some error $\sigma_c$, we can also set 95\% CL limits on  ${\rm Br}_{inv}$. For this purpose one can
use the overall PDF (from the combination of all Higgs search channels) for the signal strength parameter, which we again approximate by a Gaussian centred at $\hat\mu_c$ with standard deviation $\sigma_c$.  
As we will interpret $\hat\mu_c <1$ as coming from a nonzero ${\rm Br}_{inv}$
we restrict now $\mu$ to the interval $(0,1)$ and normalize the combined PDF
accordingly, {\it i.e.} $\int_0^1 pdf(\mu)d\mu=1$.
Then we determine a 95\% CL interval $(\mu_{L1},\mu_{L2})$ around $\hat\mu_{c}$ such that
\be
\int_{\mu_{L1}}^{\mu_{L2}}pdf(\mu)d\mu=0.95\ ,
\ee
imposing the condition that the interval is centred at $\hat\mu_{c}$ if $\mu_{L1}>0$ and $\mu_{L2}<1$. Otherwise one fixes $\mu_{L1}=0$ or $\mu_{L2}=1$. From this interval we derive a 95\% CL allowed band for  ${\rm Br}_{inv}$ as
\be
1-\mu_{L2}<{\rm Br}_{inv}<1-\mu_{L1}\ .
\ee
One can also place a comparable bound in the context of a $\chi^2$ fit that is given by
\bea
{\rm Max} \left[1-\hat\mu_c - 2 \sigma_c,0 \right] < {\rm BR}_{inv} < {\rm Min} \left[1-\hat\mu_c + 2 \sigma_c,1 \right].
\eea
The sensitivity of this $\chi^2$-based bound is expected to be equivalent to the sensitivity to ${\rm BR}_{inv}$ in the PDF test  in the Gaussian limit.
Figure \ref{fig:BRinvLimits} shows the sensitivity band, as a function of $\hat\mu_{c}$ for the PDF test, for several values of its error $\sigma_c$: the current one ($\sigma_c=0.3$); the combined error expected when both ATLAS and CMS accumulate enough data for a $5 \sigma$ Higgs discovery per experiment  over this year ($\sigma_c=0.15$), with the 95\% CL excluded region shaded; and with a future error value down to $\sigma_c=0.05$.  As expected, if $\hat\mu_{c}$ is small this requires a large invisible width and a lower limit on ${\rm Br}_{inv}$ can be set while, if $\hat\mu_{c}$ is closer to 1, then only an upper limit on ${\rm Br}_{inv}$ 
can be derived. For intermediate values of $\hat\mu_{c}$ a ``measurement" of
${\rm Br}_{inv}$ would be possible. The plot shows that the error in the determination of the true value of ${\rm Br}_{inv}$ (along the diagonal) is approximately $2\sigma_c$.
Note that for ${\rm Br}_{inv} \rightarrow 0$ or $1$, the corresponding values of $\hat\mu_{c}$ themselves require smaller $\sigma_c$ than for moderate values of $\hat\mu_{c}$ to reach a discovery. 
Thus if a discovery is actually made in these cases, any corresponding ${\rm Br}_{inv}$ will be simultaneously more accurately known than for the
$\hat\mu_{c} \sim 0.5$ case.

\begin{figure}[tb]
\includegraphics[width=0.47\textwidth]{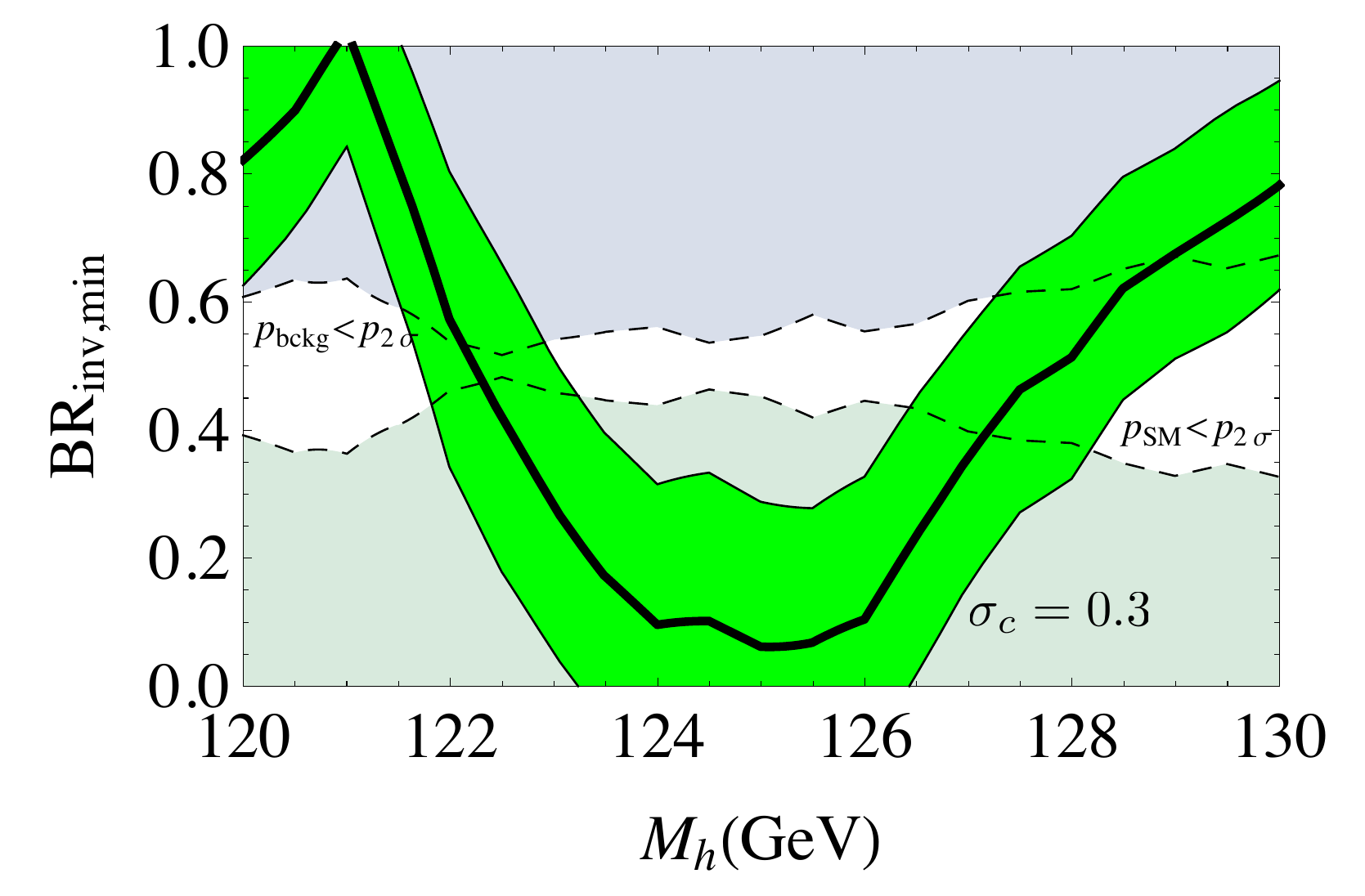}
\hspace*{0.5cm}
\includegraphics[width=0.47\textwidth]{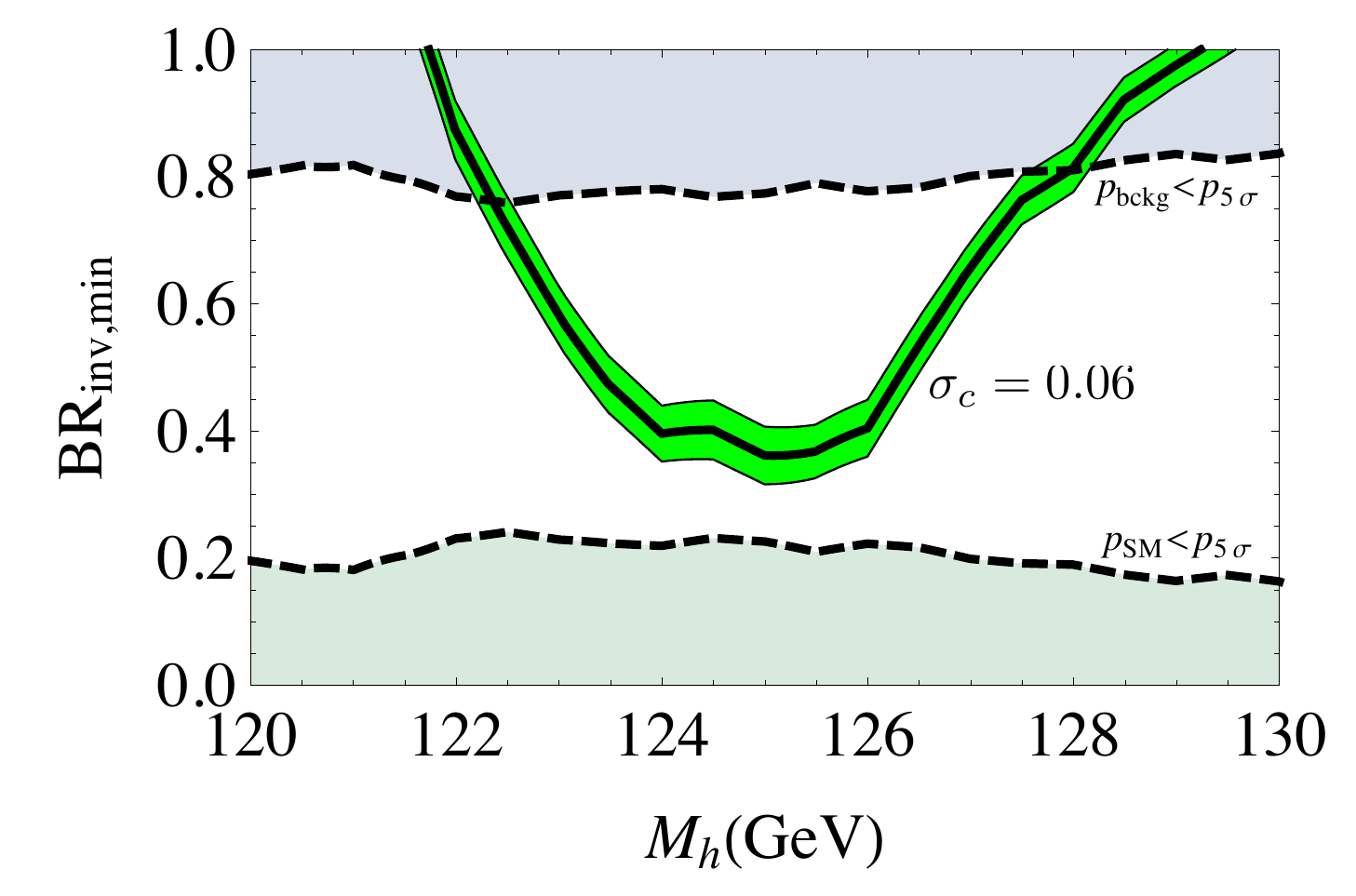}
\caption{\it Left: Current status of the experimental situation concerning ${\rm Br}_{inv}$, extracted from
combining the $\hat\mu_c$ values reported by ATLAS, CMS and the Tevatron, and interpreting deviations from $\hat\mu_c=1$ as coming from an invisible Higgs width.  Above the lower dashed line $p_{SM}<p_{2\sigma}$; below the upper dashed line, $p_{back}<p_{2\sigma}$, so that no strong evidence for a nonzero value of ${\rm Br}_{inv}$ is possible at this time.
Right: Same as left, in a hypothetical future situation (the solid curve is obtained from the left figure data series by shifting the data series to larger values of ${\rm Br}_{inv}$ and reducing the error) assuming a factor $5$ improvement in the precision with which the combined $\hat\mu_c$ could be measured compared to current data. The dashed lines correspond now to p-values equal to $p_{5\sigma}$, so that finding $5\sigma$ evidence for a nonzero ${\rm Br}_{inv}$ would be possible in the white region between both lines.}\label{fig:BRinvMh}
\end{figure}

Finally, in Fig.~\ref{fig:BRinvMh} we translate  the best-fit value $\hat\mu_c$ obtained by combining the $\hat\mu_c$  results of ATLAS, CMS and the Tevatron into a best-fit value for the ${\rm Br}_{inv}$ (using $\hat\mu_c=1-{\rm Br}_{inv}$) as a function of $m_h$.\footnote{See also Ref.~\cite{Low:2011kp} for a recent analysis (on older data) with a similar reinterpretation of the data as in Fig.~\ref{fig:BRinvMh} (left).}
The left plot shows the current situation, with a best-fit ${\rm Br}_{inv}>0$ for the interesting Higgs mass range $m_h\sim 124$ GeV, which is in any case perfectly compatible with a zero value. 
The larger values of ${\rm Br}_{inv}$ for other Higgs masses
are also not statistically significant because they correspond to values of either $p_{SM}$ or $p_{back}$ not particularly small (for reference, $p_{SM}<p_{2\sigma}$ only above the lower dashed line, while  $p_{bckg}<p_{2\sigma}$ only below the upper dashed line).
This can change with higher energy/luminosity and the right plot shows a hypothetical future situation with nonzero Higgs invisible width after collection of more data (such that the current $\sigma_c\sim 0.3$ used in the left plot is reduced by a factor 5). Besides a hypothetical curve with the best value for ${\rm Br}_{inv}$,  the plot also shows the regions of parameter space for which $p_{SM}$ and $p_{back}$ are below $5\sigma$, illustrating how such an analysis could claim indirect evidence for  ${\rm Br}_{inv} \neq 0$.

\section{Robustness of Global Fits to extract ${\bma{\rm Br}}_{\bma{inv}}$}\label{global}
In the previous section we have examined the prospects for bounding or discovering ${\rm Br}_{inv}$ for the Higgs in BSM scenarios where new physics primarily couples to the dimension two scalar mass operator.
In this section, we will examine how robust these conclusions are when a scalar resonance that has only approximately SM Higgs properties is
involved in EWSB.  First we will consider in Section \ref{acfits} the case of a minimal effective chiral EW Lagrangian with a non-linear realization of  $\rm
SU(2)_L \times U(1)_Y$ and a light scalar resonance. This scenario is most easily interpreted in composite Higgs scenarios and introduces parameters $(a,c)$ for the unknown
coupling of a scalar resonance to the gauge and fermion fields of the SM, with the SM case corresponding to $(a=1,c=1)$. (See Refs. \cite{Giudice:2007fh,Contino:2010mh,Grober:2010yv}).
The obvious problem one faces in this case is how to determine if a universal reduction in signal yields is due to a nonzero ${\rm Br}_{inv}$ or to a uniform reduction of the Higgs couplings 
involved in the search channels ({\it{i.e.}}, $a=c<1$).


The robustness of global fits for ${\rm Br}_{inv}$ in the presence of unknown 
higher dimensional operators is also important to determine. In Section \ref{SMHiggs1}, when considering the effects of  ${\rm Br}_{inv}$ on the SM Higgs, it was assumed
that all new light states are essentially not charged under $\rm SU(3)_c \times SU(2)_L \times U(1)_Y$, so that only Higgs decay (but not Higgs production) was affected, due to invisible decays to those states.
When this assumption is relaxed, one can consider BSM scenarios in which, besides the light 
SM singlets leading to ${\rm Br}_{inv}$, heavier new states [charged under $\rm SU(3)_c \times SU(2)_L \times U(1)_Y$] leave their trace in the low-energy effective theory through higher dimensional operators. In that case, it is important to examine the interplay of the effects of such higher dimensional operators on Higgs production and the effect of the light singlet states on ${\rm Br}_{inv}$ in global fits. 
We study this question in Section \ref{higherD}.

\subsection{Global Fits to a Non-SM scalar resonance and extracting ${\bma{\rm Br}}_{\bma{inv}}$ }\label{acfits}
In this section, we consider the more general case, consistent with the current data set, of a minimal effective chiral EW Lagrangian with a non-linear realization of  $\rm
SU(2) \times U(1)_Y$ and a light scalar resonance, denoted as $h$. Such a theory includes the Goldstone bosons associated with
the breaking of the weakly gauged $\rm SU(2)_L \times U(1)_Y$ (which is a subgroup of $\rm SU(2)_L \times \rm SU(2)_R$) and the  SM field content. This theory is the minimal effective theory of a light scalar degree of freedom
that can have the experimentally supported pattern of MFV flavour breaking as in the SM, and respect custodial symmetry -- $\rm SU(2)_c$, while the $W$ and $Z$ are massive. (See Ref.~\cite{Farina:2012ea} for an analysis of the LHC data relaxing the $\rm SU(2)_c$ assumption.) The Goldstone bosons 
are denoted by $\pi^a$, where $a = 1,2,3$, and are grouped as 
\bea
\Sigma(x) = e^{i \sigma_a \, \pi^a/v} \; ,
\eea
with $v = 246 \, {\rm GeV}$ and $\sigma_a$ the Pauli matrices.  The $\Sigma(x)$ field transforms linearly under $\rm SU(2)_L \times SU(2)_R$ as $\Sigma(x) \rightarrow
L \, \Sigma(x) \, R^\dagger$ where $L,R$ indicate the transformation
on the left and right under $\rm SU(2)_L$ and $\rm SU(2)_R$, respectively, while $\rm SU(2)_c$ is the diagonal subgroup of
$\rm SU(2)_L \times SU(2)_R$, under which the scalar $h$
transforms as a singlet. The leading terms in the derivative expansion of such a theory
are given by \cite{Giudice:2007fh,Contino:2010mh,Grober:2010yv,Grinstein:2007iv}
\bea
\mathcal{L} &=&  \frac{1}{2} (\partial_\mu h)^2 + \frac{v^2}{4} {\rm Tr} (D_\mu \Sigma^\dagger \, D^\mu \Sigma) \left[1 + 2 \, a \, \frac{h}{v} \right] \\
&&- \frac{v}{\sqrt{2}} \, (\bar{u}_L^i \bar{d}_L^i) \, \Sigma \, \left[1 + c \, \frac{h}{v}  \right] 
\left(
\begin{array}{c} 
y_{ij}^u \, u_R^j \\ 
y_{ij}^d \, d_R^j 
\end{array} \right)  +  h.c. + \cdots \nn
\label{eq:efflag}
\eea
Here we have neglected potential terms that are not relevant for the fits we will perform. 
Fitting the current data in such a theory has been recently explored in the literature \cite{Azatov:2012bz,Espinosa:2012ir}.
Note that the SM Higgs is a special case of this theory, and corresponds to a linear completion ($h$ becomes part of a linear multiplet) of this non-linear sigma model, with $a = c = 1$.
\subsection{Imposing EWPD}
It is useful to consider fitting to EWPD simultaneously with the global data to obtain a more constrained parameter space in this effective theory.\footnote{In fact, we will find in subsequent sections that marginalization over multiple parameters in the three dimensional fit space including EWPD is required to obtain a residual $\chi^2$ distribution that is not flat.} In EWPD analyses, the corrections to the gauge boson propagators in this effective Lagrangian
can be expressed in terms of shifts of the oblique parameters S and T \cite{Holdom:1990tc,Peskin:1991sw,Altarelli:1990zd} given by
\bea
\Delta S &\approx& \frac{-(1 - a^2)}{6 \, \pi} \, \log \left(\frac{m_h}{\Lambda}\right),  \quad \quad  \Delta T \approx \frac{3(1 - a^2)}{8 \, \pi \, \cos^2 \theta_W} \, \log \left(\frac{m_h}{\Lambda}\right).
\eea
These equations are approximate in that the numerical coefficient is determined from the logarithmic large $m_h$ dependence of $\rm S,T$ given in Ref.~\cite{Peskin:1991sw}.
Here we have introduced a Euclidean momentum cut-off scale $\Lambda$, which
approximately represents the mass of new states that are required to cut-off the growth in the longitudinal gauge boson scattering.
In a full calculation, with all degrees of freedom, the cut-off scale will cancel.  The degree to which
this Euclidean cut-off properly captures the UV regularization of these integrals by new states not included in the effective theory is model dependent.
We assume that the UV completion of the effective Lagrangian is such that directly treating this cut-off scale
as a proxy for a heavy mass scale integrated out is valid, i.e. that further arbitrary parameters rescaling the cut-off scale terms need not be introduced. 
The cut-off scale is chosen to be  $\Lambda = 4 \, \pi \, v/|\sqrt{1-a^2}|$ for $a \neq 1$.

For EWPD we use the results of the ${\it Gfitter}$ collaboration \cite{Baak:2011ze} for $m_h = 120 \, {\rm GeV}$,
\bea
S = 0.04 \pm 0.10,  \quad \quad  T = 0.05 \pm 0.11,  \quad \quad U = 0.08 \pm 0.11, \;
\eea
and the correlation coefficient matrix is given by
\bea
C = \left(
\begin{array}{ccc} 
1 & 0.89 & -0.45 \\ 
0.89 & 1 & -0.69 \\
 -0.45 & -0.69 & 1 \\
\end{array} \right).
\eea

We shift these results to having the input $m_h = 124 \, {\rm GeV}$ using the one-loop contribution of the SM Higgs field to S and T. This numerical shift is  $\lesssim 10^{-2}$.  
There is a strong preference for $a \simeq 1$ in a global fit due to EWPD, i.e the SM mechanism of mass generation of the $W^\pm$ and $Z$ is strongly preferred in minimal scenarios where EWPD can be directly interpreted to dictate the value of $a$.
When EWPD is imposed one has a bias in the fit space so that $a>1$, but this should not be over-interpreted. This bias could in principle be a hint for the existence of other
states in EWPD, but this possibility cannot be disentangled from cut-off scale effects without further experimental and theoretical input. 
We conservatively consider this bias to be simply a numerical artifact of our cut-off procedure.

Interestingly, EWPD offers a handle to disentangling the degeneracy between $a = c < 1$ and the presence of ${\rm Br}_{inv}$: $a\neq 1$ has a direct impact on EWPD, while the new singlet states (into which the Higgs can decay invisibly) can have no impact on EWPD. In any case, the possibility of such degeneracy implies that further cross-checks of  ${\rm Br}_{inv}>0$ would be needed to confirm an eventual indirect evidence coming from the global tests discussed in the last Section. 
Directly confirming such indirect evidence for ${\rm Br}_{inv}$ is best accomplished in more traditional studies of experimental sensitivity to ${\rm Br}_{inv}$
based on the kinematics of Higgs decay products. We discuss prospects for such a direct confirmation in Section \ref{direct}.

\subsection{Marginalizing/Fixing Parameters}\label{marginalizeac}
\begin{figure}[tb]
\includegraphics[width=0.31\textwidth]{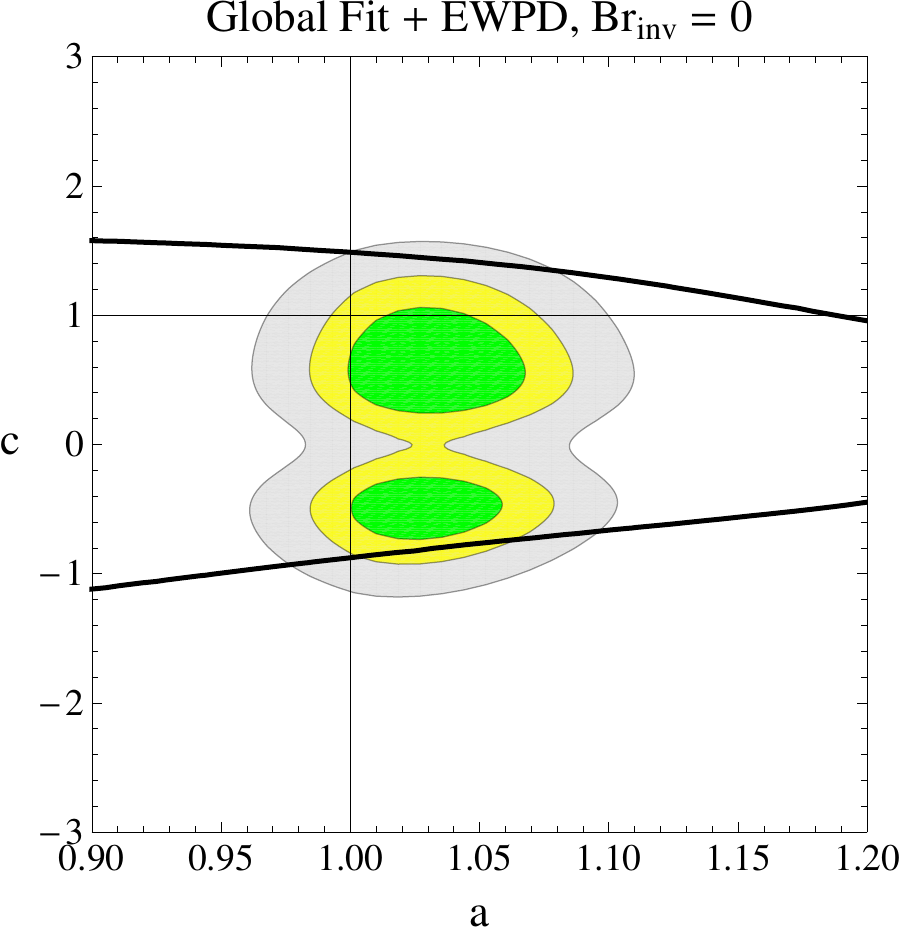}
\includegraphics[width=0.31\textwidth]{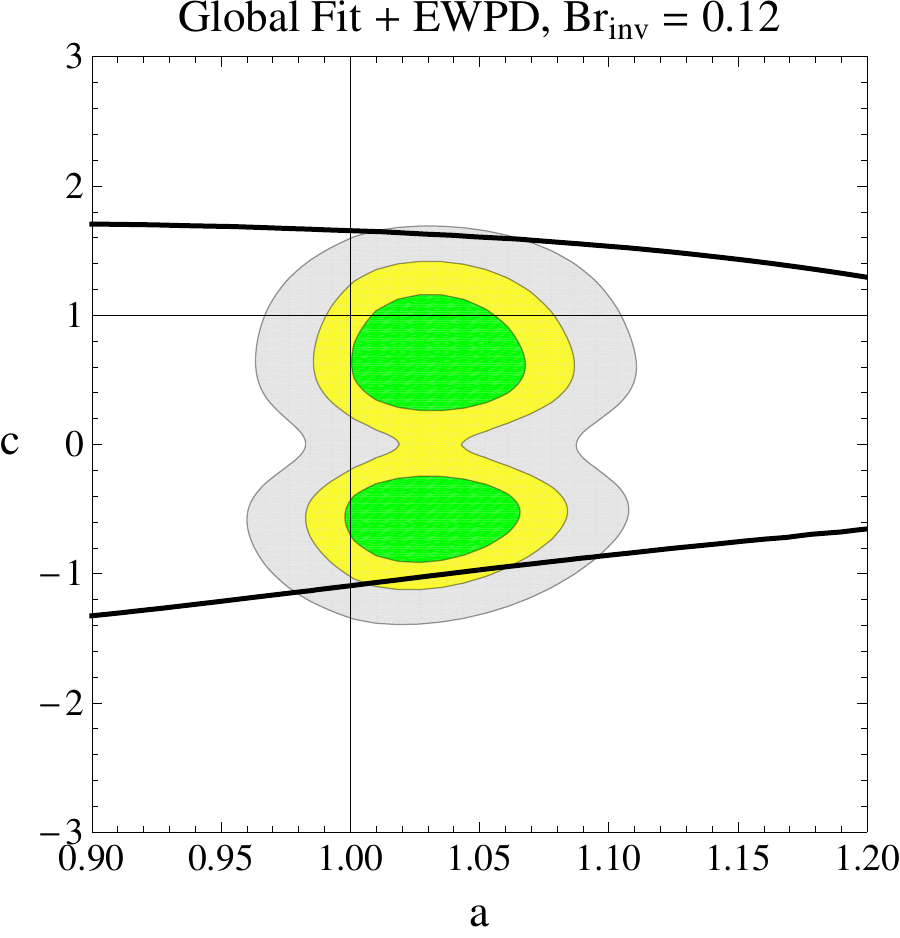}
\includegraphics[width=0.31\textwidth]{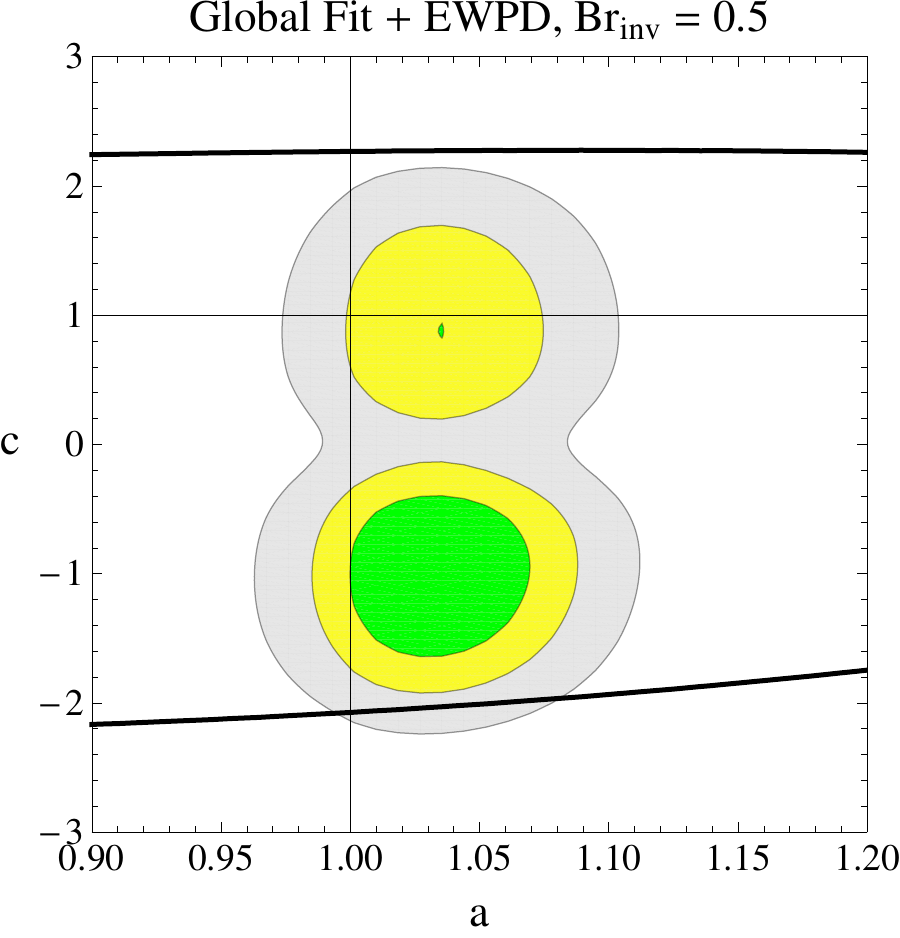}\\
\vspace{0.5cm}
\includegraphics[width=0.31\textwidth]{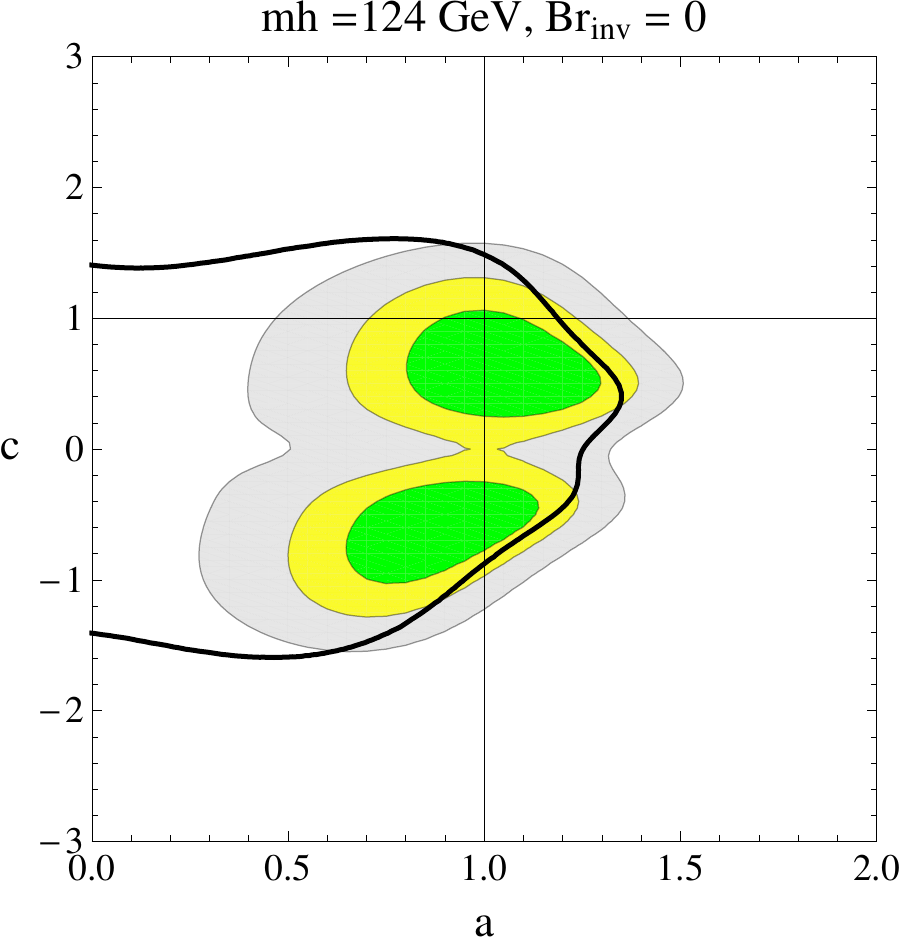}
\includegraphics[width=0.31\textwidth]{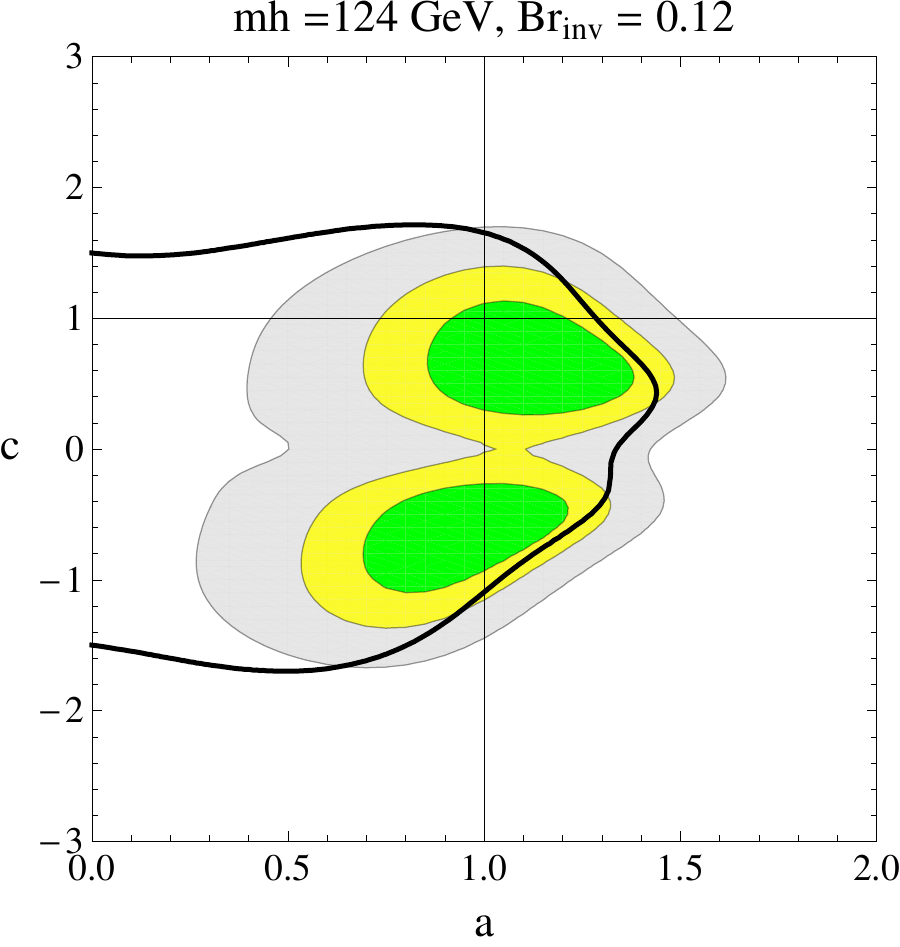}
\includegraphics[width=0.31\textwidth]{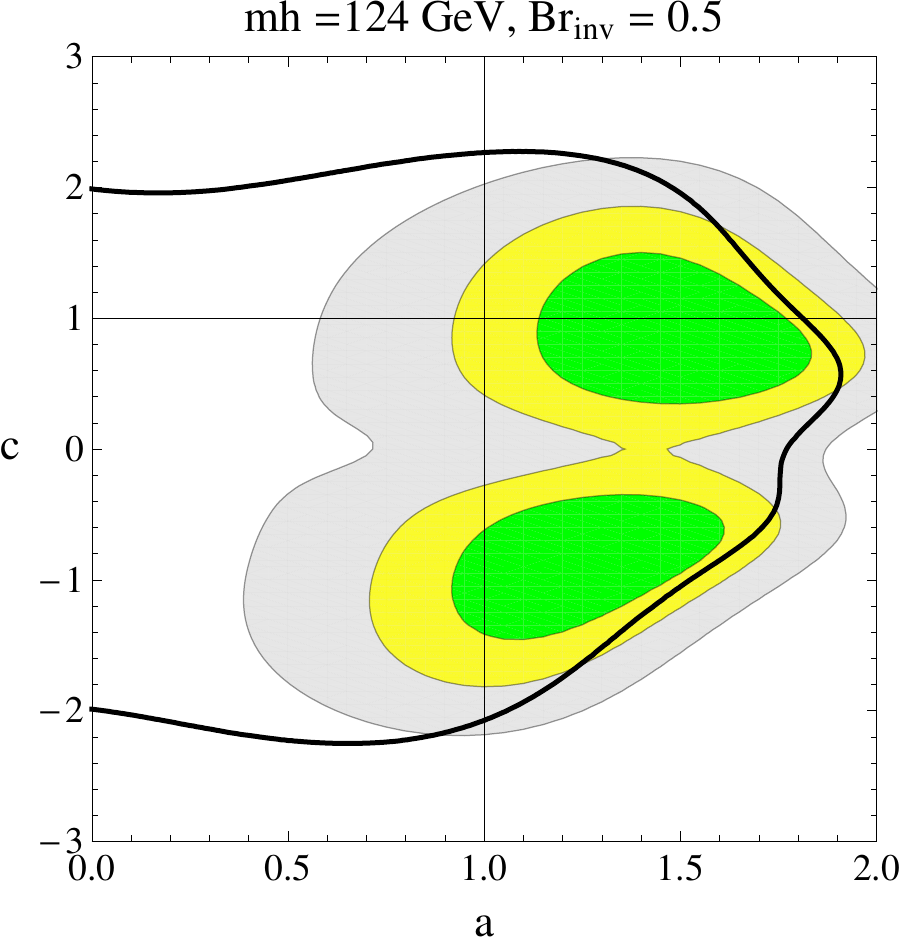}
\caption{\it Global fit to best-fit signal strengths in SM Higgs searches, for various values of ${\rm Br}_{inv}$. In the upper plots EWPD is also imposed, while in the lower plots
EWPD is not included in the global fit. Here the green region is the $65 \%$ CL region defined through the CDF for a two parameter fit. The yellow region is the $90 \%$ CL region and the grey region encloses the $99\%$ CL region. Also shown as solid black lines are the $95 \%$ exclusion regions (outside this line is excluded at $95 \%$ CL)
in the parameter space using the procedure described in Appendix B of Ref. \cite{Espinosa:2012ir} and the data in the Appendix.}\label{Fig.EWPD1}
\end{figure}
First, consider the case of fixing or marginalizing over one of the parameters $(a,c,{\rm Br}_{inv})$\footnote{In the remainder of the paper we will always choose the value $m_h = 124 \, {\rm GeV}$ as we have shown that the fit results are not strongly dependent (considering current errors) on the chosen mass (when varied in the range $124-126 \, {\rm GeV}$).}  to examine the robustness of our global fit results
when ${\rm Br}_{inv} =0$ is assumed, as in Ref. \cite{Espinosa:2012ir}. 
Fits with various ${\rm Br}_{inv}$ as an input value are shown in Fig. \ref{Fig.EWPD1}. We find that, when EWPD constraints are incorporated into the fit, the $c<0$ minimum of the $\chi^2$ is preferred for larger values of ${\rm Br}_{inv}$.  This is easy to understand, as ${\rm Br}(h \rightarrow \gamma \gamma)$
depends on the interference of fermion and gauge boson loops with an interference term $\propto - a \, c$. As the invisible width gets larger and the expected number of events
in $\gamma \gamma$ final states decreases, negative values of $c$ allow (by constructive interference) the number of $\gamma \gamma$ events to be larger and more consistent with the data which show an excess
in a number of $\gamma \gamma$ subchannels. It is interesting that this is another example where the breaking of the approximate $c \leftrightarrow -c$ symmetry in the parameter space has a physical consequence. 
If (relative) excesses in the signal strengths of $h \rightarrow \gamma \gamma$
became statistically significant with a larger data set, coincident with common suppressions in the other discovery channels, such a pattern of deviations can be explained with 
a large ${\rm Br}_{inv}$ and a negative $c$ in the effective theory, with constraints from EWPD.
\begin{figure}
\includegraphics[width=0.4\textwidth]{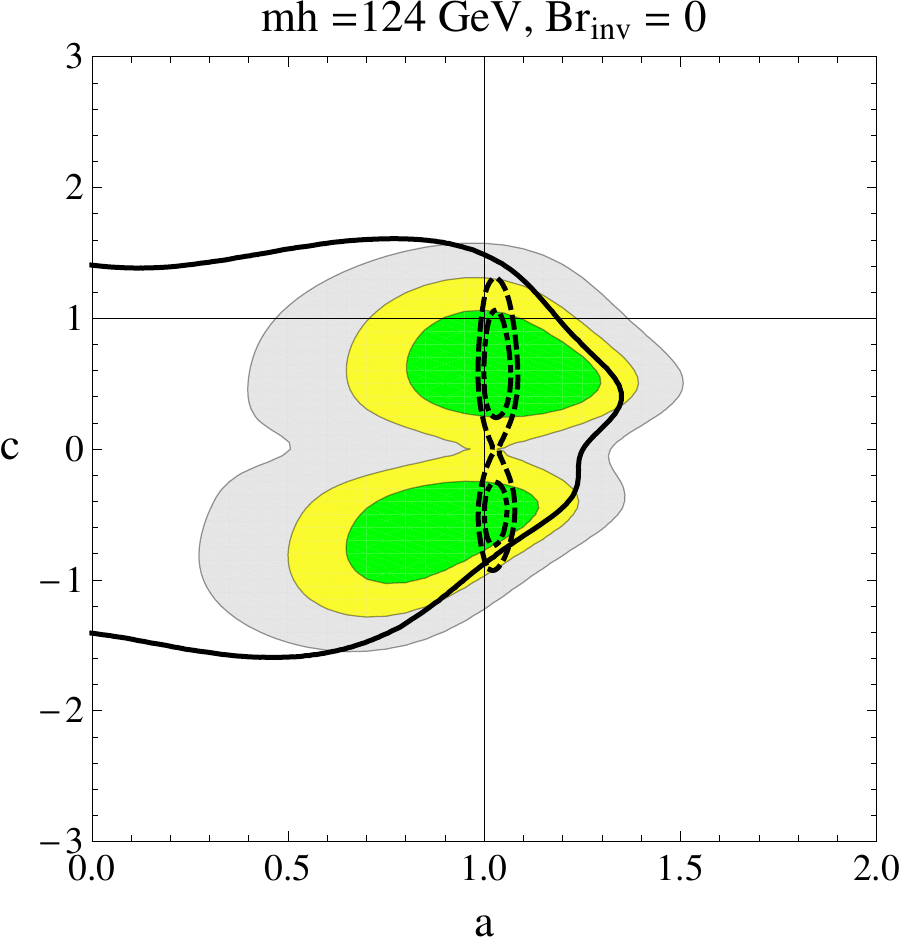} \hspace{0.5cm}
\includegraphics[width=0.42\textwidth]{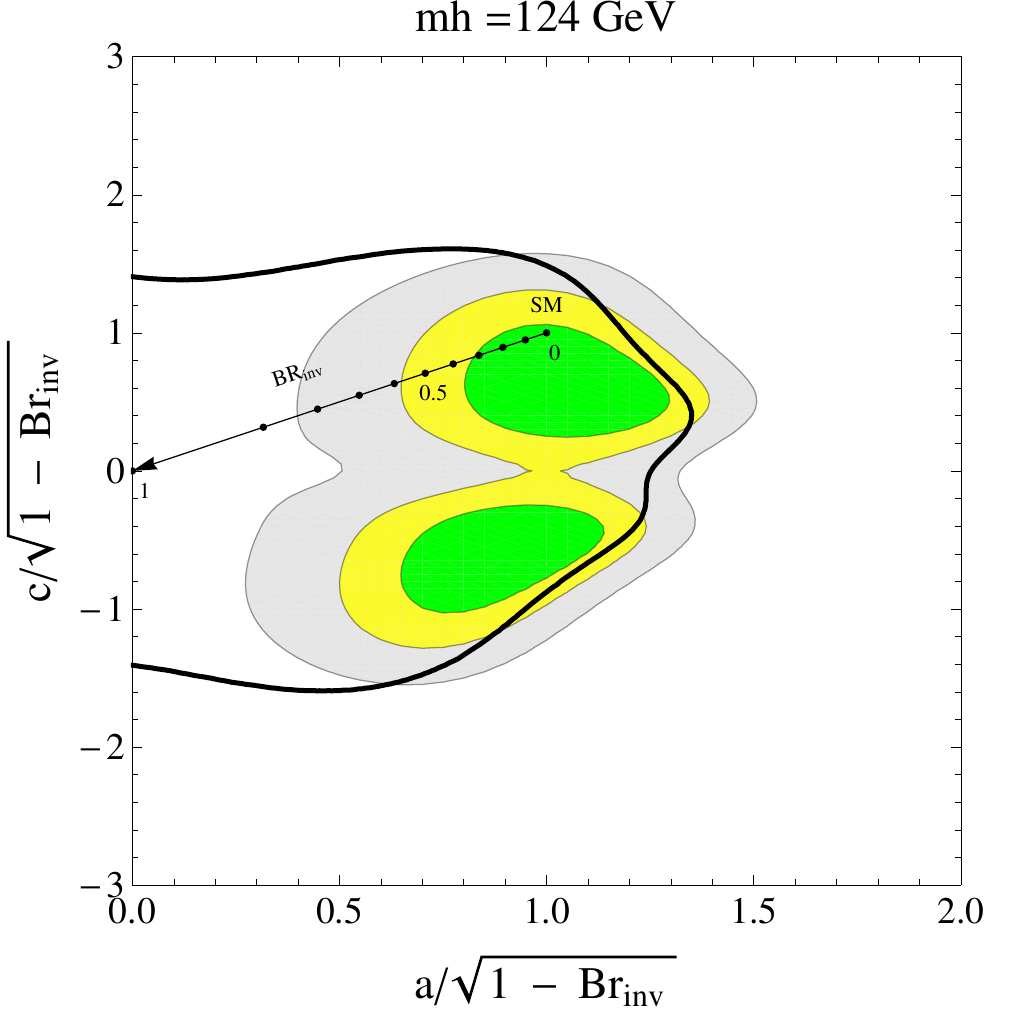}
\caption{\it Allowed parameter space of $(a,c)$ (left) comparing the fit with EWPD (dot-dashed and dashed lines for $65 \%$ and $90 \%$ CL contours) and without EWPD. 
The plot colour convention is the same as in previous figures with the solid black line again denoting the $95 \%$ CL exclusion limit. Shown on the right is the allowed parameter space in the 
scaling variables $(a/\sqrt{1-{\rm Br}_{inv}}, c/\sqrt{1-{\rm Br}_{inv}})$ with a solid line from $(1,1)$ to $(0,0)$ to mark the location of the SMinv point as a function of ${\rm Br}_{inv}$. The dots on the line represent  ${\rm Br}_{inv}$
from $0$ to $1$ in steps of $0.1$ from right to left.}\label{Fig.joses}
\end{figure}
\begin{figure}
\includegraphics[width=0.38\textwidth]{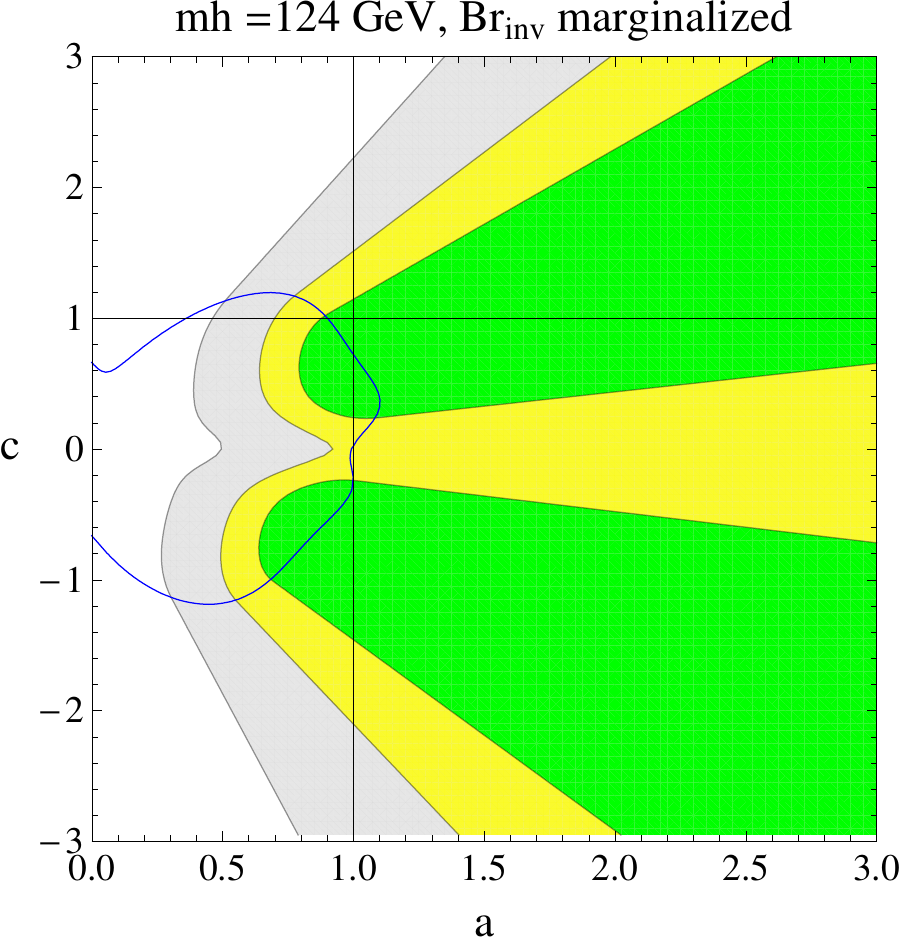} \hspace{0.5cm}
\includegraphics[width=0.38\textwidth]{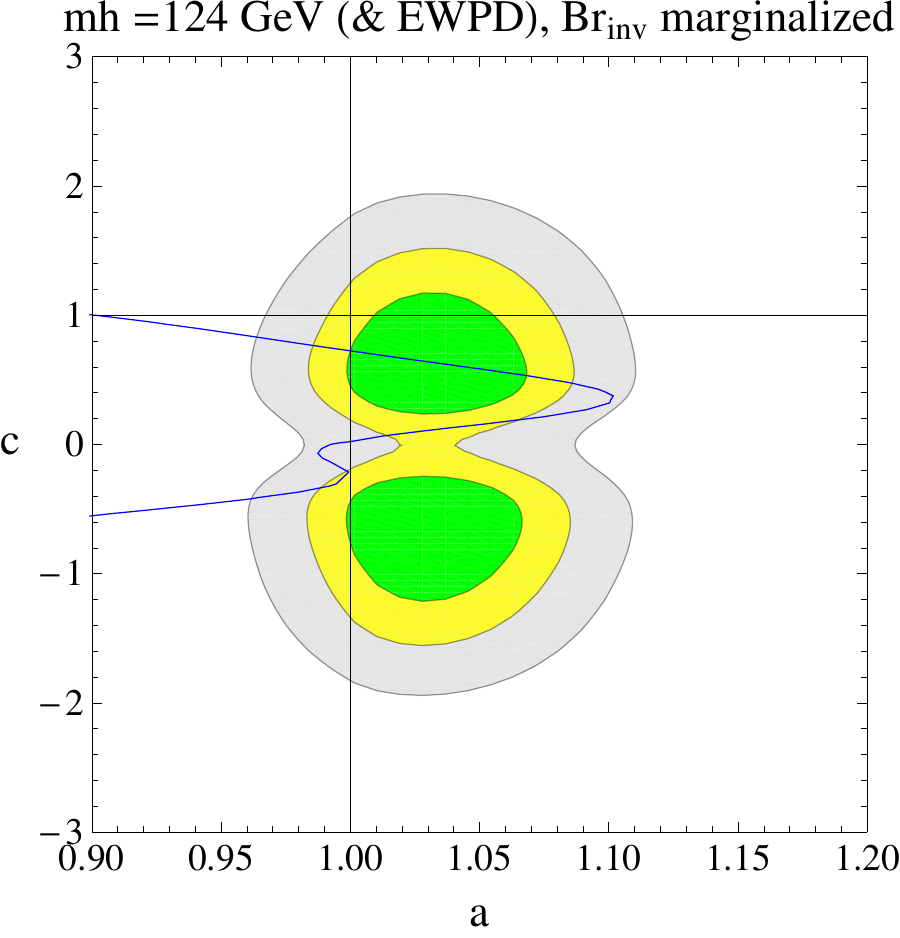}
\caption{\it Allowed parameter space of $(a,c)$ in the global fit when ${\rm Br}_{inv}$ is marginalized over, subject to the physical constraint $0 < {\rm Br}_{inv} <1$. To the left EWPD is not included in the joint fit,
for the right EWPD is also included in a global $\chi^2$. The blue solid line delimits the parameter space where the prior condition ${\rm Br}_{inv} = 0$ is satisfied (within the blue line).}\label{Fig.gammamarg}
\end{figure}
Conversely, when EWPD are not used, the allowed parameter space is shifted to larger values of
$a$ and $c$ as ${\rm Br}_{inv}$ increases to (partially) cancel the suppression of events due to ${\rm Br}_{inv}\neq 0$.
More precisely, the lower plots shown in Fig.~\ref{Fig.EWPD1} have a simple scaling property corresponding to a dilatation from the origin in $(a,c)$ space,
relating the spaces in plot $i$ to plot $j$ as
\bea\label{scaling}
(a_i,c_i) \sqrt{1-{\rm Br}_{inv}^i} = (a_j,c_j) \sqrt{1-{\rm Br}_{inv}^j}.
\eea 
The constraints from EWPD on the fit space can be more easily understood by directly comparing the fit spaces as shown in Fig.\ref{Fig.joses} (left).
Similarly, the dilatation scaling of the best fit space is illustrated in Fig.\ref{Fig.joses} (right) where we plot the fit space
as a function of the scaling variables $(a/\sqrt{1-{\rm Br}_{inv}}, c/\sqrt{1-{\rm Br}_{inv}})$.

\begin{figure}
\includegraphics[width=0.38\textwidth]{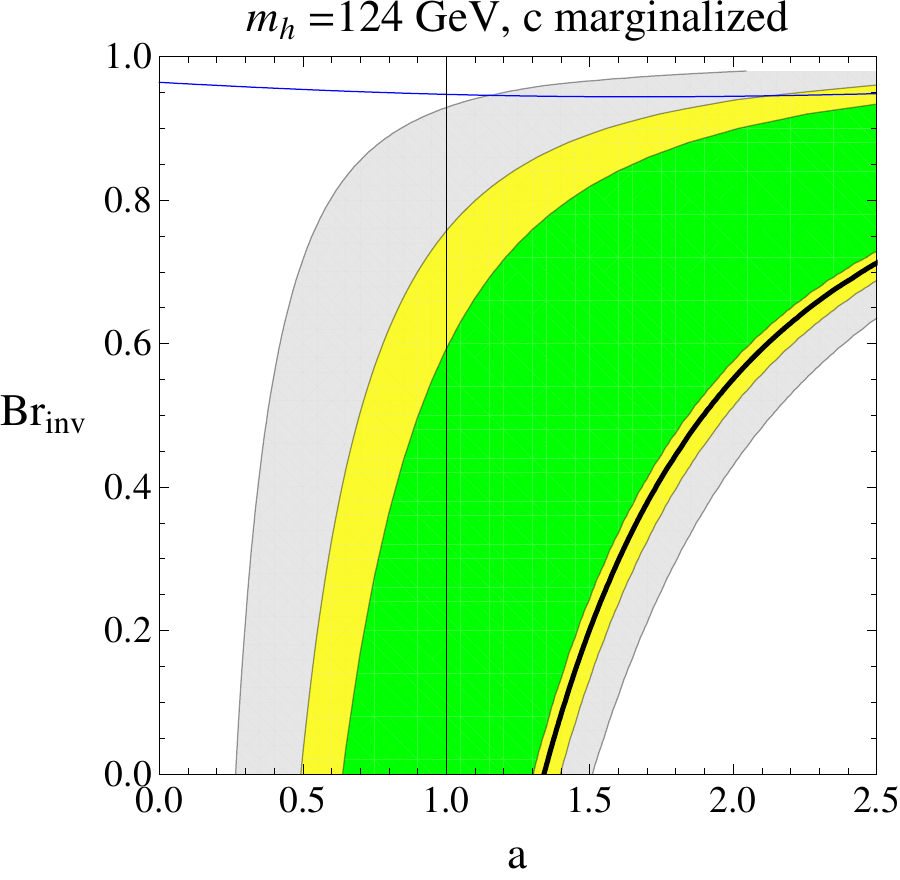}  \hspace{0.5cm}
\includegraphics[width=0.38\textwidth]{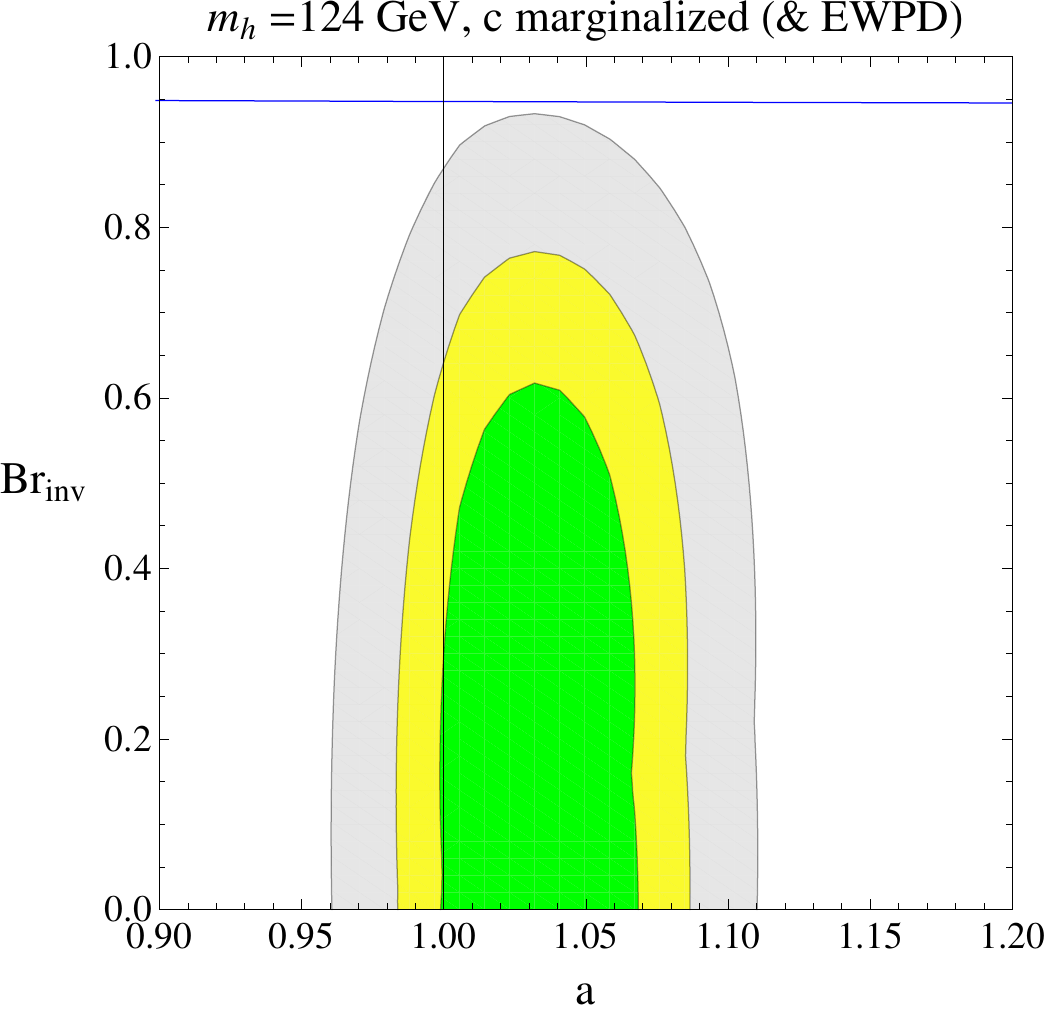}
\caption{\it Allowed parameter space of $(a,{\rm Br}_{inv})$ in the global fit when $c$ is marginalized over, subject to the constraint
$|c| < 3.5$. The solid blue line illustrates where this prior is saturated (above the blue line). Again, on the right (left) panel EWPD is (not) included in the global $\chi^2$ fit.
The plot colour convention is the same as in previous figures with the solid black line again denoting the $95 \%$ CL exclusion limit in the parameter space, to compare with the best fit regions.}\label{Fig.cmarg}
\end{figure}
\begin{figure}
\includegraphics[width=0.42\textwidth]{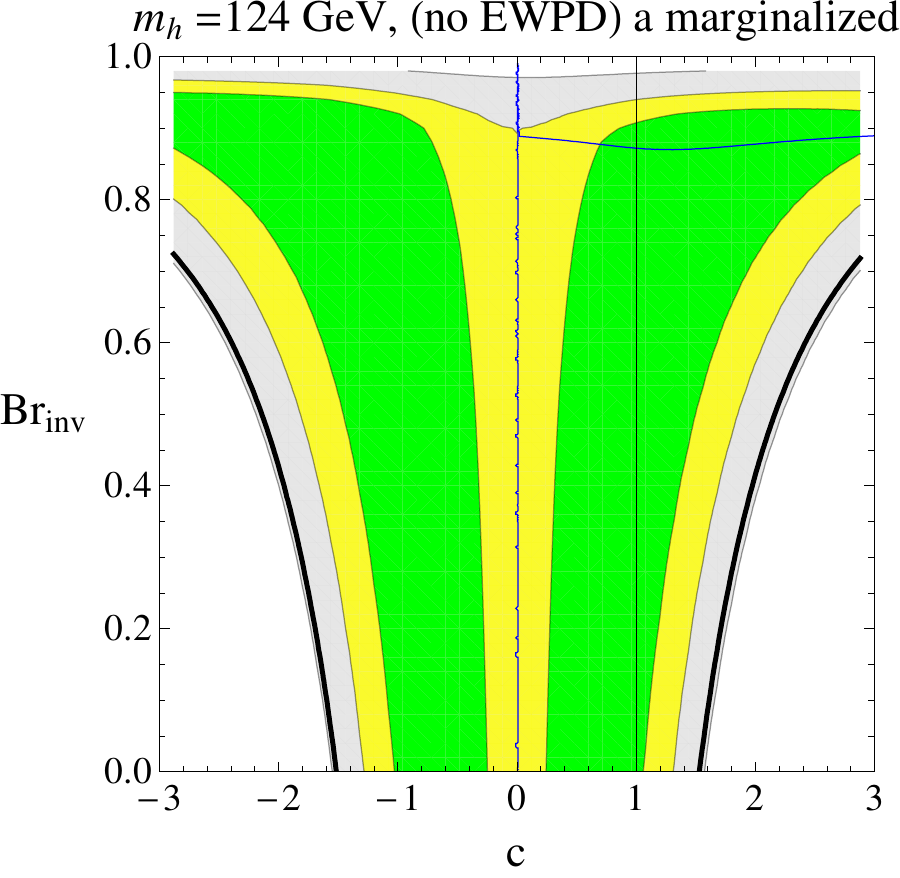}  \hspace{0.5cm}
\includegraphics[width=0.42\textwidth]{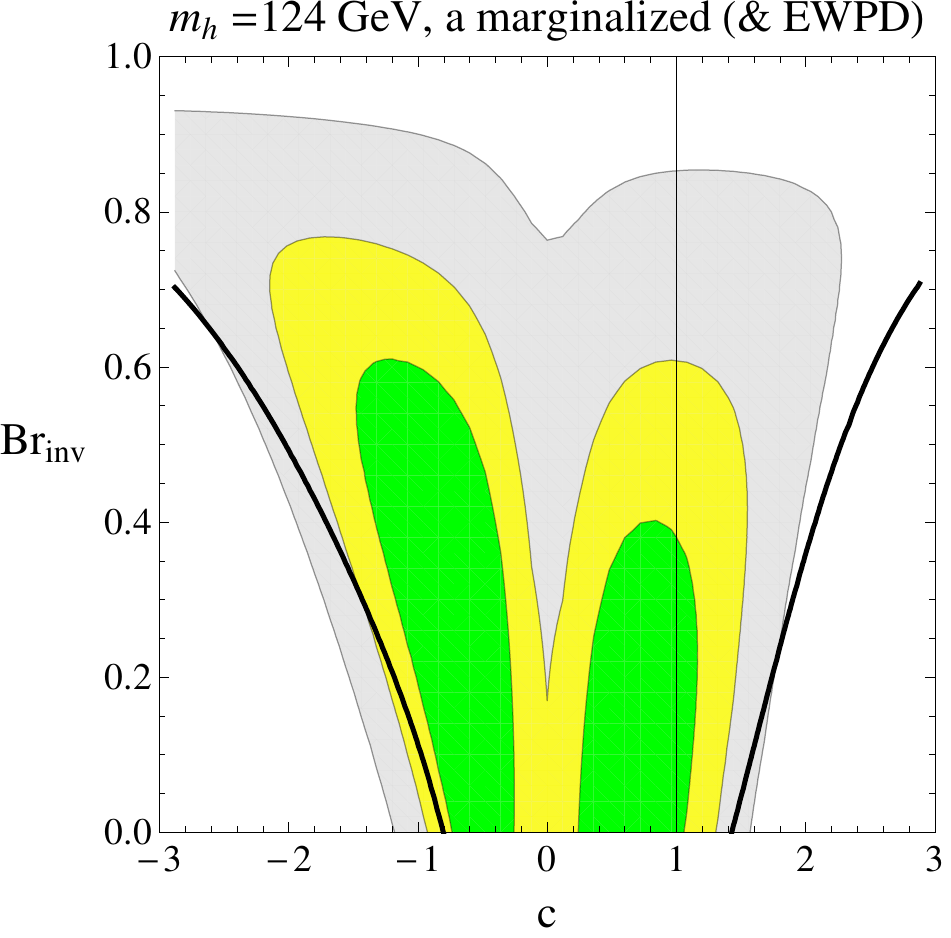}
\caption{\it Allowed parameter space of $(c,{\rm Br}_{inv})$ in the global fit when $a$ is marginalized over, subject to the constraint
$0< a < 3$. The solid blue line illustrates again where the prior is saturated: above the horizontal line, $a = 3$ while, to the left of the vertical solid blue line, $a=0$. Note that, in the right figure, the prior is never saturated as EWPD forces $a \sim 1$.
Again in the right (left) panel EWPD is (not) included in the global $\chi^2$ fit. The plot colour convention is the same as in previous figures, and the $95 \%$ CL exclusion contours are black solid lines.}\label{Fig.amarg}
\end{figure}

Now consider treating each one of the parameters $(a,c,{\rm Br}_{inv})$ as a nuisance parameter in turn. Doing so we can also examine the effects of an unknown parameter on the remaining fit space.
For example, in marginalizing over ${\rm Br}_{inv}$  we define a reduced $\chi^2$ function
\bea
\chi^2(a,c) = \chi^2(a,c,{\rm Br}_{inv}(a,c)),
\eea
where ${\rm Br}_{inv}(a,c)$ is given by the solution of $d \chi^2(a,c,{\rm Br}_{inv})/d {\rm Br}_{inv} = 0$. Then the allowed parameter space is defined through the CDF for a two parameter fit, and we 
obtain the results in Fig. \ref{Fig.gammamarg}. Marginalizing over the parameters $c$ and $a$ we find the results shown in  Fig. \ref{Fig.cmarg} and Fig. \ref{Fig.amarg} respectively which demonstrate the
correlation between the allowed ${\rm Br}_{inv}$, and the allowed parameter space for the remaining unknown parameters. This correlation is due to the dilatation relationship shown in Eq.~\ref{scaling}.

\begin{figure}[t]
\includegraphics[width=0.45\textwidth]{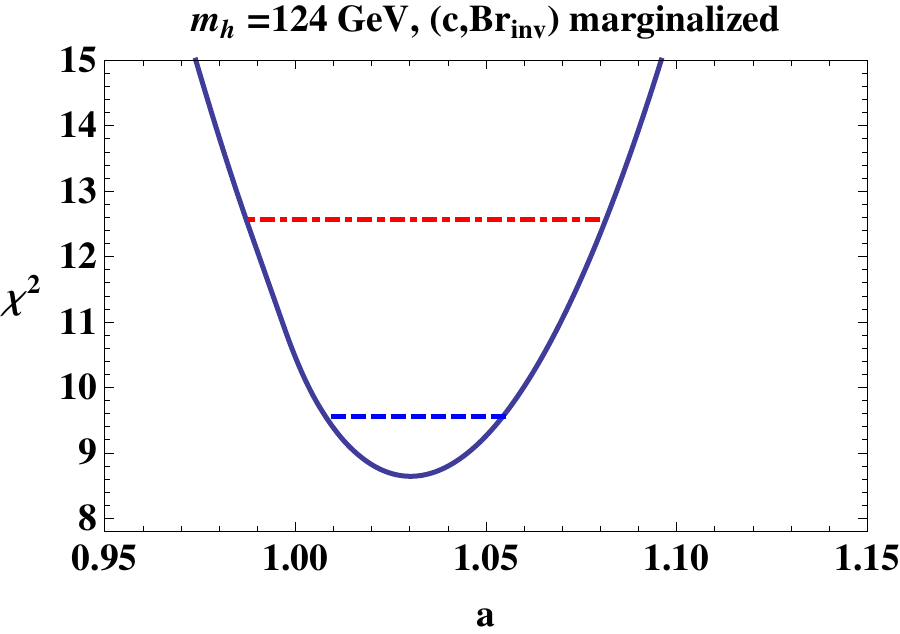}  \hspace{0.5cm}
\includegraphics[width=0.44\textwidth]{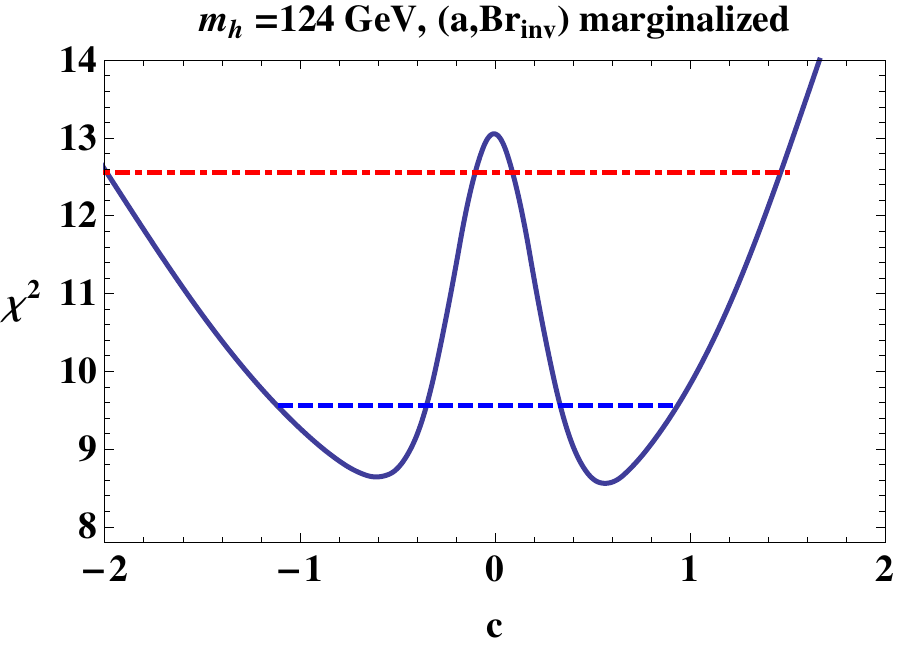}
\includegraphics[width=0.45\textwidth]{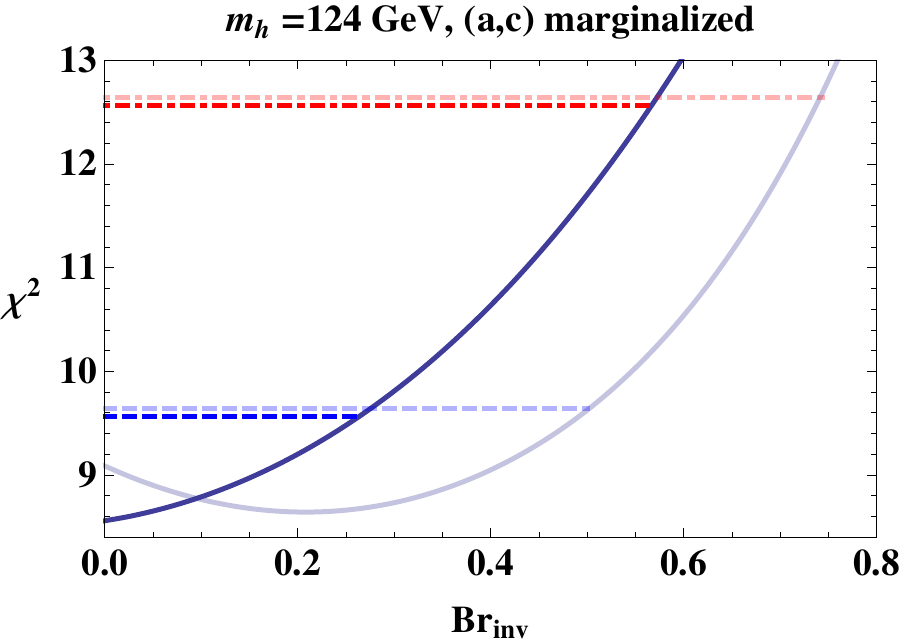}
\caption{\it Marginalizing over the two dimensional space to obtain the $\chi^2$ distribution for a single parameter. EWPD is included in the global $\chi^2$ measure. The blue dashed horizontal line in each
plot is the $68 \%$ CL ($1 \sigma$) limit, the red dot-dashed line in each figure is the $95 \%$ CL ($2 \sigma$) limit. The CL regions are defined by the cumulative distribution function for a one parameter fit.
The bottom figure shows the two curves from the nearly degenerate $\chi^2_{min}$ with $c>0$ (unfaded lines for $(\chi^2_{min})_1$) or $c<0$ (faded lines for $(\chi^2_{min})_2$). 
The difference in the minima is negligible $[(\chi^2_{min})_2 - (\chi^2_{min})_1]/(\chi^2_{min})_1 = 0.01$ with $(\chi^2_{min})_1$ slightly prefered.
When the $c<0$ minima is chosen, and the marginalization is performed, the preference for ${\rm Br}_{inv}$ in the current data increases due to the interference effects previously discussed.}\label{Fig.doublemarg}
\end{figure}

Finally one can marginalize two of the free parameters simultaneously in order to obtain the residual $\chi^2$ distribution to examine if the slight statistical preference for 
${\rm Br}_{inv} >0$ persists. In this case, one
must impose EWPD to avoid a flat distribution in the remaining free parameter. We find the results shown in  Fig. \ref{Fig.doublemarg}. 
Of most interest is the result of marginalizing over free gauge and fermion couplings, while imposing EWPD. In this case, 
one finds that the global fit in this theory is now ${\rm Br}_{inv} = 0$, with the $95 \%$ CL limit ${\rm Br}_{inv} < 0.57$, for a scalar mass of $124 \, {\rm GeV}$.
Comparing this to the SMinv result of Section II, we see that an unknown $a,c$ can remove the slight preference for  ${\rm Br}_{inv} = 0$ in the current global fits when the $\chi^2_{min}$ with $c>0$ is the global minimum. Conversely when 
the $\chi^2_{min}$ with $c<0$ is chosen, the slight preference for ${\rm Br}_{inv}$ in the data set is not removed. Then the best fit is ${\rm Br}_{inv} = 0.21$, with the $95 \%$ CL limit ${\rm Br}_{inv} < 0.75$.
This result makes clear that performing such a two dimensional marginalization over $(a,c)$
is an important cross check to see if any evidence of ${\rm Br}_{inv} \neq 0$ is robustly preferred in a future data set.

\subsection{Higher Dimensional Operators, ${\bma{\rm Br}}_{\bma{inv}}$, and the SM Higgs.}\label{higherD}

It is also of interest to consider the impact of possible BSM states on Higgs production and decay, when evidence for ${\rm Br}_{inv} >0$ emerges from global fits. 
The exact impact of BSM states on Higgs phenomenology is model dependent.
In this section, we consider the case where new states that are
SM singlets lead to ${\rm Br}_{inv}$, and other new states, that are charged under $\rm SU(3)_c \times SU(2)_L \times U(1)_Y$, or at least a subgroup of the SM group, lead to higher dimensional operators. 
Our aim is to examine the degree to which conclusions about ${\rm Br}_{inv}$ can be extracted from global fits
in the context of unknown Wilson coefficients of the resulting higher dimensional operators.

Assuming that these BSM states do not source CP violation,
the operators of interest for Higgs phenomenology (in global fits to $\hat{\mu}_i$) are given by
\bea
\mathcal{L}_{HD} &=& - \frac{c_G \, g_3^2}{2 \, \Lambda^2} \, H^\dagger \, H \, G^A_{\mu\, \nu} G^{A \, \mu \, \nu} - \frac{c_W \, g_2^2}{2 \, \Lambda^2} \, H^\dagger \, H \, W^a_{\mu\, \nu} W^{a \, \mu \, \nu} 
 - \frac{c_B \, g_1^2}{2 \, \Lambda^2} \, H^\dagger \, H \, B_{\mu\, \nu} B^{\mu \, \nu}, \nn \\
&\,&   - \frac{c_{WB} \, g_1 \, g_2}{2 \, \Lambda^2} \, H^\dagger \, \tau^a \, H \, B_{\mu\, \nu} W^{a \, \mu \, \nu}. 
\eea
Note that $g_1,g_2,g_3$ are the weak hypercharge, $\rm {SU}(2)$ gauge and $\rm{SU}(3)$ gauge couplings and we are using the notation of 
Ref. \cite{Manohar:2006gz}.
The scale $\Lambda$ corresponds to the mass scale of the lightest new state that is integrated out.
We are primarily interested in the effects on $\sigma_{gg \rightarrow h}$ and $\Gamma_{h \rightarrow \gamma \gamma}$ as these are loop level processes in the SM, sensitive to BSM
effects. As we expect loop level
contributions to these operators from the BSM states, we rescale the Wilson coefficients as $c_j = \tilde{c}_j/(16 \pi^2)$ for $j = G,W,B,WB$ and fit to combinations of $\tilde{c}_j$.

Using the results of Ref. \cite{Manohar:2006gz}, the effect of these operators are 
\bea\label{higher-d.effect}
\sigma_{gg \rightarrow h} \approx \sigma^{SM}_{gg \rightarrow h} \, \left| 1 - (1.39  + 0.10 \, i) \frac{v^2 \, \tilde{c}_{G}}{\Lambda^2}\right|^2,   \quad \! \!
\Gamma_{h \rightarrow \gamma \, \gamma} \approx \Gamma^{SM}_{h \rightarrow \gamma \, \gamma} \, \left| 1 + 0.15 \frac{v^2 \, \tilde{c}_{\gamma}}{\Lambda^2}\right|^2.  
\eea
Here $\tilde{c}_{\gamma} = \tilde{c}_W + \tilde{c}_B - \tilde{c}_{WB}$ and we have used $m_t = 172.5 \, {\rm GeV}$, $m_h = 124 \, {\rm GeV}$ and $\alpha_s(172.5) = 0.1095$.
The imaginary part of the numerical coefficients above comes from including the $b$ quark loop correction (we use $m_b = 4.7 \, {\rm GeV}$) in normalizing the BSM effect to the SM amplitudes. Normalizing in this manner is done
to reduce the SM dependence in the BSM correction when this rescaling is used in our fits, and a numerical value is used for  $\sigma^{SM}_{gg \rightarrow h}$.
As in Ref. \cite{Manohar:2006gz},  we have retained the two loop QCD correction to the SM matching of the $h \, G^A_{\mu\, \nu} G^{A \, \mu \, \nu} $ operator in the $m_t \rightarrow \infty$ limit in these numerical coefficients. Due to this choice, 
this  correction cancels out (in the $m_t \rightarrow \infty$ limit) of the overall coefficient of the BSM effects when multiplied by the numerical value of $\sigma^{SM}_{gg \rightarrow h}$. This is a $\sim 10 \%$ correction on the quoted numerical coefficient. Initial state radiation
and vertex corrections to $G^A_{\mu\, \nu} G^{A \, \mu \, \nu}$ are expected to be common multiplicative factors for the operator $h \, G^A_{\mu\, \nu} G^{A \, \mu \, \nu} $ in the $m_t \rightarrow \infty$ limit,
and as such are not incorporated in the numerical factors multiplying $\tilde{c}_{G}, \tilde{c}_{\gamma}$ above.
We will consider the parameter space where the SM is modified by these corrections and  ${\rm Br}_{inv} \neq 0$ in this section, fitting to $(v^2 \, \tilde{c}_{\gamma}/\Lambda^2,v^2 \,\tilde{c}_{G}/\Lambda^2,{\rm Br}_{inv})$.
The exact relationship between these parameters, if any,  is model dependent and unknown. As such, we fit to the data assuming no relationship between the three parameters.\footnote{See also Refs.~\cite{Carmi:2012yp,Giardino:2012ww,Batell:2011pz}, for example, for recent fits to BSM higher dimensional operator Wilson coefficients based on Higgs signal strength parameters.}
  
The operators in $\mathcal{L}_{HD}$ also affect ${\rm Br}(h \rightarrow \gamma \, Z)$, where a different combination of the Wilson coefficients $\tilde{c}_W, \tilde{c}_B,\tilde{c}_{WB}$ enters.
This branching ratio is subdominant to the ${\rm Br}(h \rightarrow \gamma \, \gamma)$ branching ratio. (Numerically the values are
${\rm Br}(h \rightarrow \gamma \, \gamma) = 2.29 \times10^{-3}$, and ${\rm Br}(h \rightarrow \gamma \, Z) = 1.46 \times10^{-3}$. However, recall that when looking for the Higgs, 
the $Z$ decay has to be multiplied by ${\rm Br}(Z \rightarrow \ell \, \ell)$.)
We neglect these effects when fitting for the allowed parameter space.
We also do not include the effects of these operators on $h \rightarrow W \, W, Z \, Z$ as the SM contribution is tree level for these processes.
Further, we also neglect effects due to higher dimensional operators possibly modifying the differential distributions of the Higgs decay products, indirectly affecting the $\hat{\mu}_i$ through modifying 
the effective signal efficiency for specific kinematic cuts. Such effects are expected to be negligible compared to the current uncertainties.
However, we do not neglect the rescaling effect on $\Gamma_{h \rightarrow gg}$ that is identified with the rescaling on $\sigma_{gg \rightarrow h}$ in Eq.~\ref{higher-d.effect}.
We include this rescaling consistently, which has a non-negligible impact  on all branching ratios through the modification of $\Gamma_{\rm SM}$.

\begin{figure}[tb]
\includegraphics[width=0.32\textwidth]{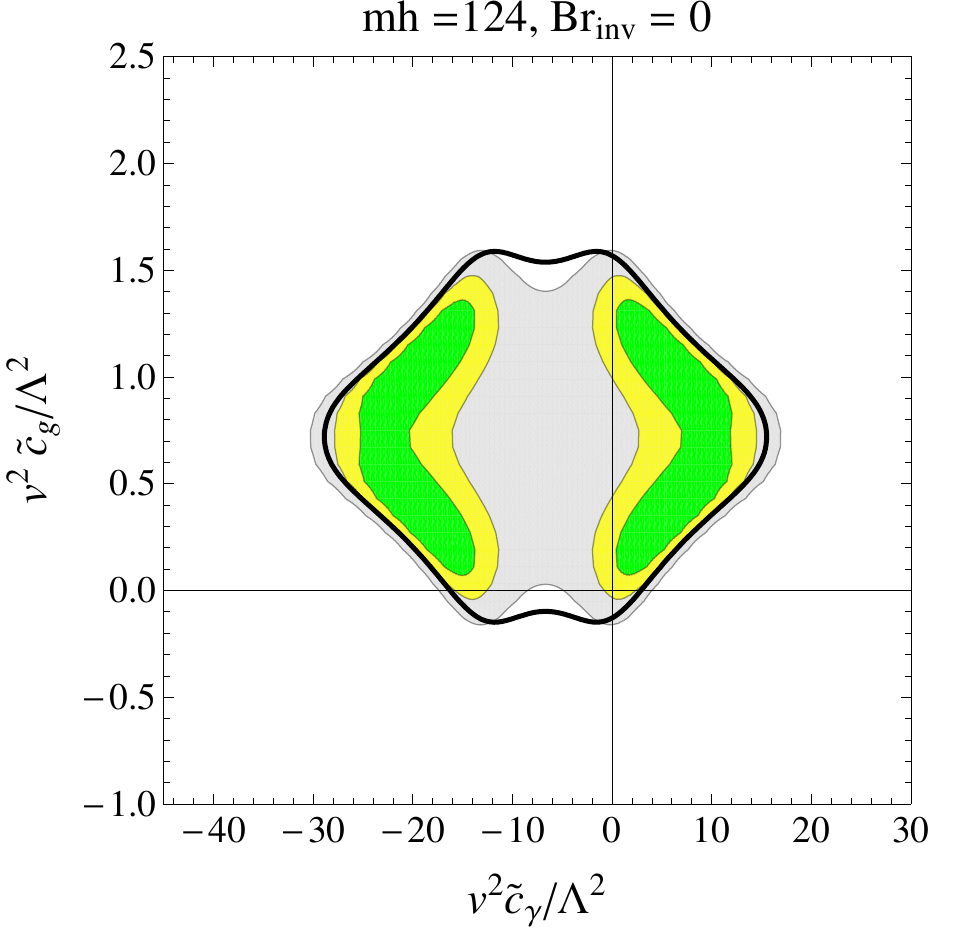}
\includegraphics[width=0.32\textwidth]{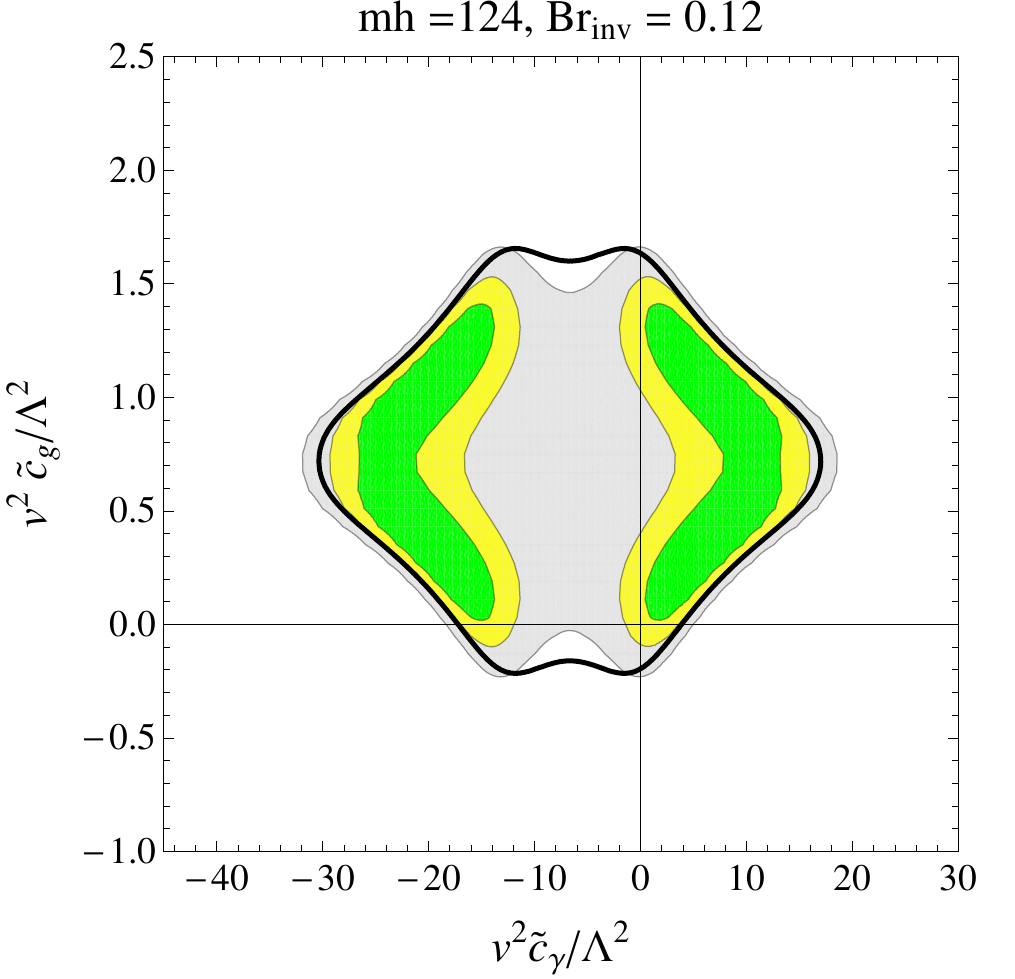}
\includegraphics[width=0.32\textwidth]{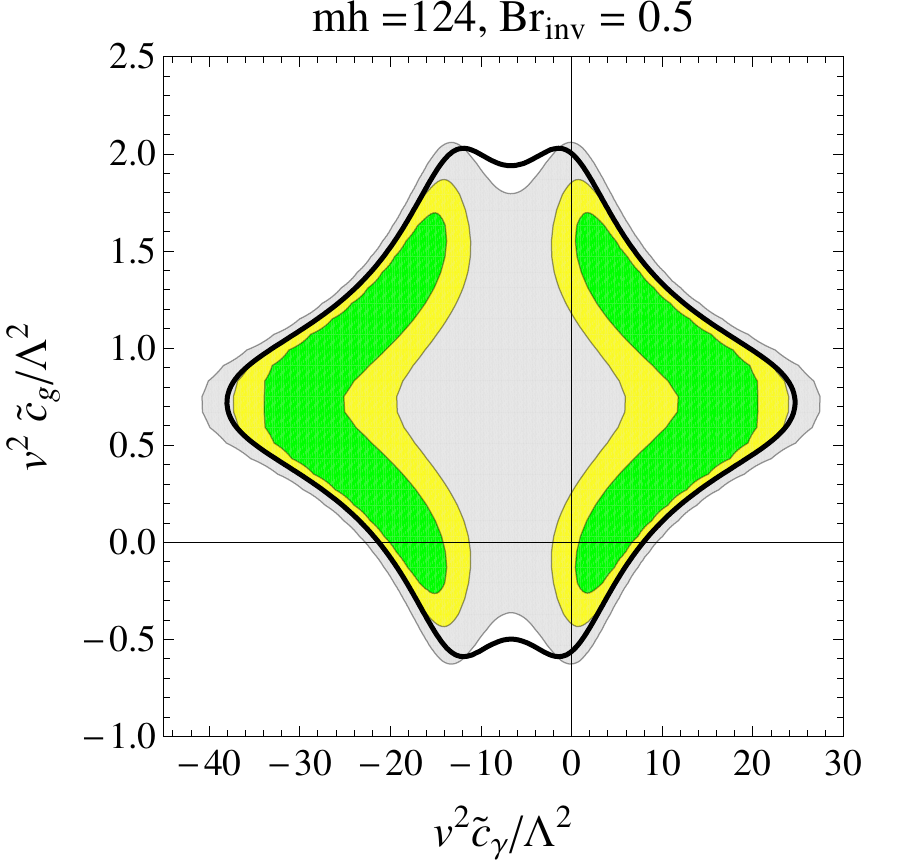}
\caption{\it Global fits to  $\tilde{c}_{G}, \tilde{c}_{\gamma}$  assuming a SM Higgs for ${\rm Br}_{inv}$ fixed to $(0,0.12,0.5)$. EWPD is not simultaneously imposed.
The convention for the plot regions is the same as previous figures with the black line delineating the $95 \%$ CL exclusion contour. In examining the allowed parameter space recall that the factor $1/16 \pi^2$ has been scaled out of the BSM contribution,
so that large allowed values of $v^2 \, \tilde{c}_{\gamma}/\Lambda^2,v^2 \,\tilde{c}_{G}/\Lambda^2$, although difficult to model build, are still perturbative corrections to the SM.}\label{Fig.HD1}
\end{figure}

\begin{figure}[tb]
\includegraphics[width=0.4\textwidth]{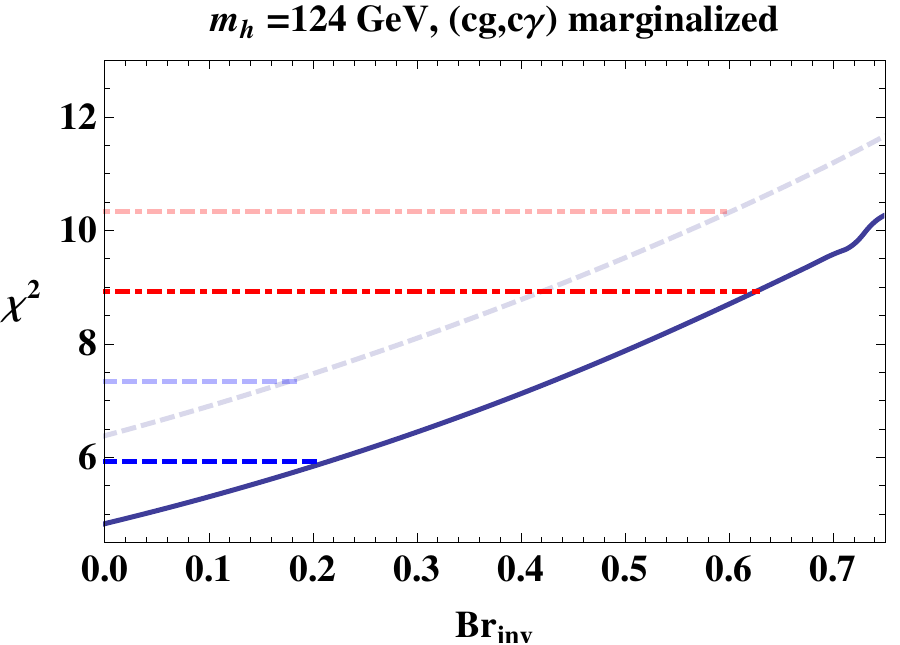}\hspace{1cm}
\includegraphics[width=0.4\textwidth]{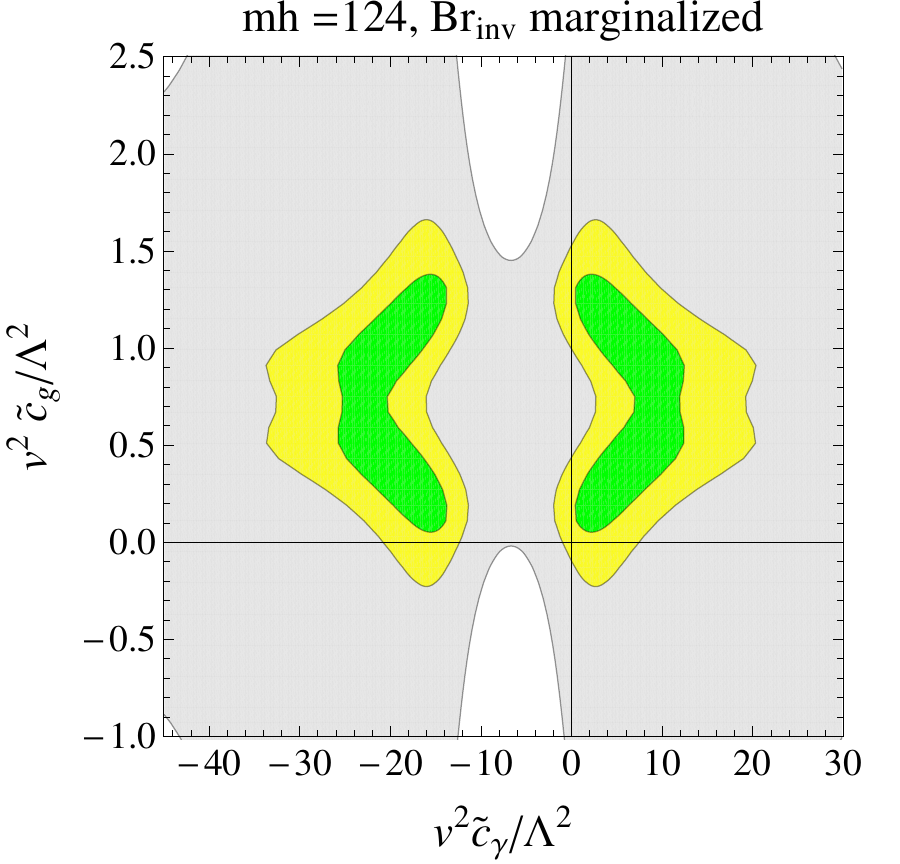}
\caption{\it Marginalizing over  $\tilde{c}_{G}, \tilde{c}_{\gamma}$  in the global fit (assuming a SM Higgs) and constructing the residual $\chi^2$ distribution for ${\rm Br}_{inv}$ (left figure).
We show the results of two marginalization procedures in this figure. The solid (unfaded) line corresponds to marginalizing over  $\tilde{c}_{G}, \tilde{c}_{\gamma}$ without any prior condition on these Wilson coefficients
imposed. Conversely, the solid faded line corresponds to marginalizing over  $\tilde{c}_{G}, \tilde{c}_{\gamma}$ with the prior that the absolute values of each of these Wilson coefficients is $<1$.
For comparison, the right figure shows the result of marginalizing over ${\rm Br}_{inv}$ when fitting for $\tilde{c}_{G}, \tilde{c}_{\gamma}$ in the current global fit. The plot colour scheme conventions are the same
as in previous figures.}\label{Fig.HD2}
\end{figure}

We show in Fig.\ref{Fig.HD1} the allowed parameter space when ${\rm Br}_{inv}$ is fixed to prior values of $0$, $0.12$ or $0.5$.
In Fig.\ref{Fig.HD2} the residual $\chi^2$ distribution for ${\rm Br}_{inv}$ is shown when $\tilde{c}_{G}, \tilde{c}_{\gamma}$  are marginalized over (left)
and we also show the allowed $\tilde{c}_{G}, \tilde{c}_{\gamma}$ parameter space when  ${\rm Br}_{inv}$ is marginalized over (right) subject to the prior constraint $0 \leq  {\rm Br}_{inv} \leq 1$.
These results show the significant impact of the higher dimensional operators, in scenarios consistent with the assumptions of this section, on attempts to extract ${\rm Br}_{inv}$ from global fits to Higgs signal strength data.

Most notably, we find that the slight preference in the global $\chi^2$ distribution for  ${\rm Br}_{inv}>0$ is removed when marginalizing over such unknown BSM effects in the current data set. This
offers further caution to over interpreting the slight preference in the global $\chi^2$ distribution for  ${\rm Br}_{inv}>0$ at this time. Although in Fig.\ref{Fig.HD1} the required Wilson coefficient $\tilde{c}_{\gamma}$ to still obtain
a good fit when ${\rm Br}_{inv} \gg 0$ is large, we find that even restricting  $\tilde{c}_G,\tilde{c}_{\gamma}$ to clearly perturbative couplings ($\leq 1$), expected in many models,
the preference for a ${\rm Br}_{inv}>0$ is removed. This result is shown in Fig.\ref{Fig.HD2} (left). 

\section{Prospects for Direct confirmation of  ${\bma{\rm Br}}_{\bma{inv}}$.}\label{direct}
As stated above, there is a degeneracy between the case $a=c<1$ and
the existence of a non-zero invisible decay. The former leaves the
branching ratios unchanged due to a common suppression factor in the
couplings, while all production channels are suppressed by the same common
factor. On the other hand, in the simple case of leaving the SM
couplings unchanged but allowing for an invisible width, the production
channels are unchanged and the branching ratios are affected as in
Eq.~\ref{eq:brinv}, leading to a common overall suppression of
production times branching ratio compared to the SM. If a signal is
seen with suppressed event rate with respect to the SM expectation the
degeneracy between these two cases 
can only be removed by observing directly a non-vanishing
invisible decay. 

It has been shown that associated production with gauge
bosons, weak boson fusion and associated production with top
quarks allows one to discover a Higgs boson decaying
invisibly, and to probe the invisible branching ratio. The typical
signature is large missing transverse energy/momentum. Assuming an
invisible 
branching ratio of 1, a Higgs boson with mass up to about 150~GeV can
be discovered in Higgs radiation from a $Z$ boson at $\int {\cal L}=
10$~fb$^{-1}$ and $\sqrt{s}=14$~TeV
\cite{Choudhury:1993hv,Frederiksen:1994me,Godbole:2003it,Davoudiasl:2004aj,Zhu:2005hv,Gagnon,Meisel,Warsinsky}. At
high luminosity this reach can be extended to $\sim 250$~GeV in
associated production with a top quark pair
\cite{Warsinsky,Gunion:1993jf,Kersevan:2002zj}. Weak boson fusion
allows for the discovery up to 480 GeV with 10 fb$^{-1}$ integrated
luminosity
\cite{Warsinsky,Eboli:2000ze,Cavalli:2002vs,Girolamo}. Assuming SM
production, invisible
branching ratios as low as 25\% 
can be probed in weak boson fusion for a 120 GeV Higgs boson
at $\int {\cal L}=30$~fb$^{-1}$ and $\sqrt{s}=14$ TeV at 95\% CL
\cite{Eboli:2000ze,Cavalli:2002vs,Girolamo}. In associated production 
with a $Z$ boson, branching ratios down to 45\% can be probed
\cite{Godbole:2003it,Meisel,Gagnon} while associated 
Higgs production with a top quark pair probes invisible branching
ratios down to 56\% \cite{Kersevan:2002zj}. 

Recent papers have investigated
the potential of a $7$~TeV collider in direct searches for an invisible
Higgs boson
\cite{Bai:2011wz,Englert:2011us,Englert:2011aa,Djouadi:2012zc,Frigerio:2012uc}. The
invisible branching ratio of a 125 GeV Higgs boson produced in weak boson
fusion with SM strength can be constrained down to $\sim 40\%$ at
$\sqrt{s}=7$ TeV and $\int {\cal L}= 20$~fb$^{-1}$
\cite{Bai:2011wz}. Monojet searches from CMS based on 4.7~fb$^{-1}$
\cite{cmsmonojet}
constrain $\xi = \sigma/\sigma_{SM} BR(h\to \mbox{inv})$ down to 1.3
at 95\% CL translating to the constrained value of the invisible branching
ratio in case of SM couplings \cite{Djouadi:2012zc}.

\begin{figure}
\includegraphics[width=0.5\textwidth]{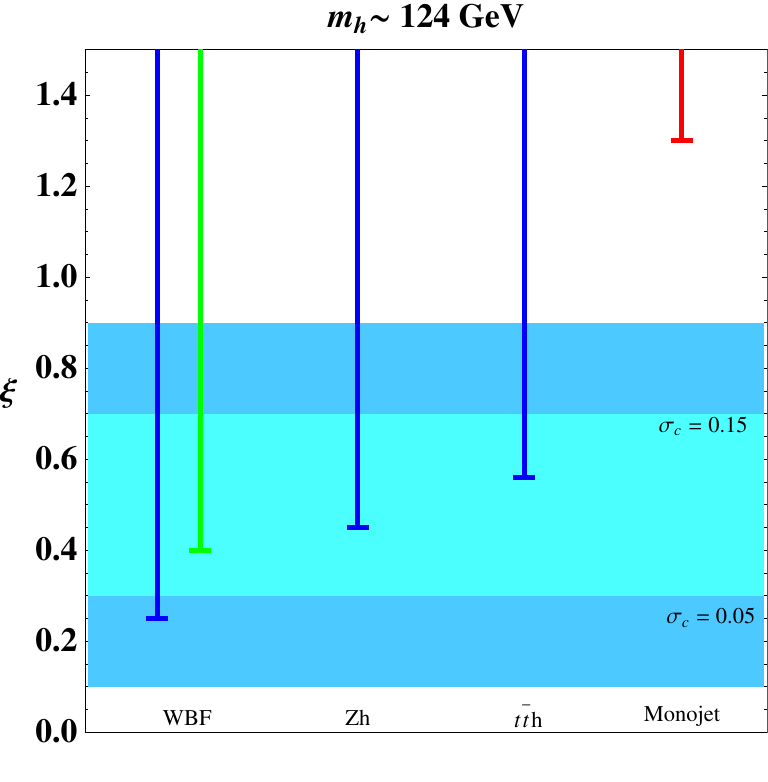}
\caption{\it The sensitivity of different analyses of direct invisible Higgs decays can be assessed by comparing their potential in setting a 95\% CL upper limit on $\xi = \sigma/\sigma_{SM} BR(h\to  \mbox{inv})$.  In the SM, with only ${\rm Br}_{inv}$ added as a free parameter, one identifies $\xi \equiv {\rm Br}_{inv}$. The vertical lines show
this reach for the indicated analyses (see labels at the bottom of the figure).  Also shown is the expected $2 \, \sigma$ sensitivity band for the SM (where again $\xi \equiv {\rm Br}_{inv}$) using the global PDF test we advocate to compare with the sensitivity of these 
  searches. The sensitivity band corresponds to the p-value test shown in Fig.\ref{Fig.scBRinv} (right). The light blue shaded horizontal band is for $\sigma_c = 0.15$, expected at the end of this year, and dark blue shaded horizontal band for $\sigma_c = 0.05$. Blue vertical lines:
  $\sqrt{s}=14$~TeV, $\int{\cal L}=30$~fb$^{-1}$, in weak boson
  fusion, $Zh$ and $t\bar{t}h$ production (from Ref.~\cite{Warsinsky}). Green vertical line: $\sqrt{s}=7$~TeV, $\int{\cal L} =
  20$~fb$^{-1}$, in weak boson fusion \cite{Bai:2011wz}. Red vertical line:
  $\sqrt{s}=7$~TeV, $\int{\cal L} = 20$~fb$^{-1}$, in monojet searches
  \cite{Djouadi:2012zc}. 
  }\label{Fig.brinvdirect} 
\end{figure}

The claimed $95 \%$ CL limits on
$\xi$ expected for $m_h \approx 124$~GeV from the direct searches discussed are summarized in
Fig.~\ref{Fig.brinvdirect}. In this figure we also show for a direct comparison the $2 \sigma$ sensitivity
expected in the global test statistics we advocate for the end of this year, with $\sigma_c = 0.15$ and
a future value when $\sigma_c = 0.05$.  One sees in this figure that the global test statistics are likely
to offer a significantly improved reach for the existence of ${\rm Br}_{inv}$ in the data set after the 2012 run.
However, to claim a discovery of  ${\rm Br}_{inv}$ will  require a combination of these global searches and direct
kinematic searches, as we have demonstrated throughout Section \ref{global}. 

\section{Conclusions.}\label{concl}
In this paper, we have systematically examined the potential of global fits
to extract information on ${\rm Br}_{inv}$ in the present and future signal strength data sets of a scalar resonance.
We have focused on the case of a scalar resonance with mass  $\sim 124 \, {\rm GeV}$, and have performed a global fit to the SM Higgs using the current signal strength data set, demonstrating that 
current global fits find the $95 \%$ CL limit ${\rm Br}_{inv} < 0.64$ for $m_h = 124 \, {\rm GeV}$. We have also illustrated how these results are statistically limited at this time
and that any statistically significant conclusion on best fit values of ${\rm Br}_{inv}$ will require more data.
We have developed a new approach to globally combining signal strengths using global PDF's to optimize searches for new states that  couple to the SM
through the `Higgs portal'. These promising results have lead us to examine the ability of global fits to resolve information on  ${\rm Br}_{inv}$  in the presence of unknown
new physics effects simultaneously impacting the properties of the Higgs. Although disentangling these effects would require further experimental input, our results make clear the
correlations expected between interpreting a global fit as providing evidence of ${\rm Br}_{inv}$ and the sensitivity of such claims to  other (unknown) new physics effects in the scenarios we have considered. 
Although the current signal strength data set we have considered in our numerical investigations only offers marginal evidence for a scalar resonance (and its properties) with $m_h\sim 124-126$ GeV, the
correlations with other new physics effects and the tests for evidence of ${\rm Br}_{inv}$ we have explored are of continued interest as the data set evolves.
\appendix

\section{Data Used}

The data we have used in the global fits of this paper are summarized in the table below.
Due to an apparent inconsistency in the ATLAS best fit signal strength plot for $ h \rightarrow b \, \bar{b}$ and the corresponding ATLAS $\rm CL_s$ limit plot 
(that is under investigation by ATLAS) we do not use the $b \, \bar{b}$ best fit signal strength value in the combined fit at this time. 
For the $pp \rightarrow  \gamma \, \gamma  \, jj $ signal of CMS we assume a $3 \%$ contamination due to $gg$ Higgs production events so that
the relevant signal rate is given by
\bea
\left(0.03 \, \sigma_{gg\rightarrow h} + \sigma_{jjh} \right) \times {\rm Br}(h \rightarrow \gamma \, \gamma).
\eea
Here $ \sigma_{jjh}$ is given by $\rm VBF$ Higgs production.
We do not use sub classes of $\rm WW$ events due to the lack of experimentally reported contaminations of these signal strengths due to other Higgs production processes.
Simultaneously using a global best fit value $\hat{\mu}$ for $\gamma \, \gamma $
events (for example) while also using a best fit $\hat{\mu}$ for a subclass of events, such as $\gamma \, \gamma  \, jj $
can result in a double counting of signal strengths that would incorrectly bias the fit. We avoid such double counting in our use of CMS and ATLAS data as the photon classes we use are exclusive,
but note that double counting of this form is present in Ref. \cite{Giardino:2012ww}, making it difficult to compare results.
In particular, to avoid introducing such a bias is why we use the experimentally reported global ATLAS $\hat{\mu}_{\gamma \gamma}$, as
a complete set of subchannel di-photon signal strengths is not available (in contrast to CMS). This is also the reason that 
we do not simultaneously use constructed signal strengths $\hat{\mu}_{\gamma \gamma}$ and $\hat{\mu}_{\gamma \gamma, P_T>40 {\rm Gev}}$ (from fermiophobic \cite{fermiophobic} searches).
These signal strengths are not independent mutually exclusive event classes, being derived from the same signal event data.
Our approach to this issue is different than the approach of Ref. \cite{Giardino:2012ww}.
\begin{table}[h] 
\setlength{\tabcolsep}{5pt}
\center
\begin{tabular}{c|c|c} 
\hline \hline 
Channel [Exp] &  $\hat{\mu}_{124}$  & $ \hat{\mu}_{125}$ 
\\
\hline
$p\bar{p} \rightarrow W^+ \, W^- \,\, [{\rm CDF \& D 0\! \! \! /}]$ \,\, Ref.\cite{TEVNPH:2012ab} & $0.35^{+1.08}_{-0.31}$  & $0.03 ^{+1.22 }_{-0.03}$  
\\
$p\bar{p} \rightarrow  b \, \bar{b} \,\, [{\rm CDF \& D 0\! \! \! /}]$ \,\, Ref.\cite{TEVNPH:2012ab} & $1.9^{+0.8}_{-0.6}$  & $2.0^{+0.8}_{-0.7}$ 
\\
$pp \rightarrow  \tau \, \bar{\tau} \,\, [{\rm ATLAS}]$  \, \, Ref.\cite{moriond}  & $-0.1^{+1.7}_{-1.7}$ & $0.1^{+1.7}_{-1.8}$ 
\\
$pp \rightarrow  Z \, Z^\star \rightarrow \ell^+ \, \ell^- \, \ell^+ \, \ell^- \,\, [{\rm ATLAS}] $  \, \, Ref.\cite{ATLASdoc2011163} & $1.6^{+1.4}_{-0.8}$ &  $1.4^{+1.3}_{-0.8}$  
\\
$pp \rightarrow  W \, W^\star \rightarrow \ell^+ \, \nu  \, \ell^- \, \bar{\nu} \,\, [{\rm ATLAS}] $  \, \, Ref.\cite{moriond} & $0.1^{+0.7}_{-0.7}$ &  $0.1^{+0.7}_{-0.6}$ 
\\
$pp \rightarrow \gamma \, \gamma \,\, [{\rm ATLAS}]$ \,\, Ref.\cite{ATLAS:2012ad} & $0.8^{+0.8}_{-0.7}  $ & $1.6^{+0.9}_{-0.8} $ 
\\
$pp \rightarrow  \tau \, \bar{\tau} \,\, [{\rm CMS}]$ \, \,  Ref.\cite{Chatrchyan:2012tx} & $0.4^{+1.0}_{-1.2}$  & $0.6^{+1.1}_{-1.2}$ 
\\
$pp \rightarrow  b \, \bar{b} \,\, [{\rm CMS}] $ \, \,  Ref.\cite{Chatrchyan:2012tx} & $1.2^{+1.9}_{-1.8}$ & $1.2^{+2.1}_{-1.8}$ 
\\
$pp \rightarrow  Z \, Z^\star \rightarrow \ell^+ \, \ell^- \, \ell^+ \, \ell^- \,\, [{\rm CMS}] $ \, \,  Ref.\cite{Chatrchyan:2012tx} & $0.5^{+1.1}_{-0.7}$& $0.6^{+0.9}_{-0.6}$ 
\\
$pp \rightarrow  W \, W^\star \rightarrow \ell^+ \, \nu  \, \ell^- \, \bar{\nu} \,\, [{\rm CMS}] $  \, \,  Ref.\cite{Chatrchyan:2012tx} & $0.6^{+0.7}_{-0.7}$  & $0.4^{+0.6}_{-0.6}$ 
\\
$pp \rightarrow  \gamma \, \gamma \,\, [{\rm CMS}] $, \, Cat.4/BDT3, \, \, Refs.\cite{Chatrchyan:2012tw,CMSdoc12001} &  $4.1^{+4.6}_{-4.1}$ & $0.6^{+1.8}_{-1.8}$
\\
$pp \rightarrow  \gamma \, \gamma \,\, [{\rm CMS}] $, \, Cat.3/BDT2, \, \, Refs.\cite{Chatrchyan:2012tw,CMSdoc12001} & $0.0^{+2.9}$ &  $2.2^{+1.5}_{-1.4}$
\\
$pp \rightarrow  \gamma \, \gamma \,\, [{\rm CMS}] $, \, Cat.2//BDT1, \, \, Refs.\cite{Chatrchyan:2012tw,CMSdoc12001} &  $2.1^{+1.5}_{-1.4}$ &$ 0.6^{+1.0}_{-0.9}$
\\
$pp \rightarrow  \gamma \, \gamma \,\, [{\rm CMS}]$, \, Cat.1/BDT0, \, \, Refs.\cite{Chatrchyan:2012tw,CMSdoc12001} &  $1.5^{+1.1}_{-1.0}$ & $2.1^{+2.0}_{-1.6}$
\\
$pp \rightarrow  \gamma \, \gamma  \, jj \,\, [{\rm CMS}]$ \,\, Refs.\cite{Chatrchyan:2012tw,CMSdoc12001} & $3.7^{+2.5}_{-1.8}$ & $3.6^{+2.2}_{-1.6}$
\\
\hline \hline
\end{tabular}
\caption{\it Summary table of reported best fit signal strengths for various Higgs mass values.}
\label{table:tcorrections} \vspace{-0.35cm}
\end{table}
\begin{figure}[ht]
\includegraphics[width=0.48\textwidth]{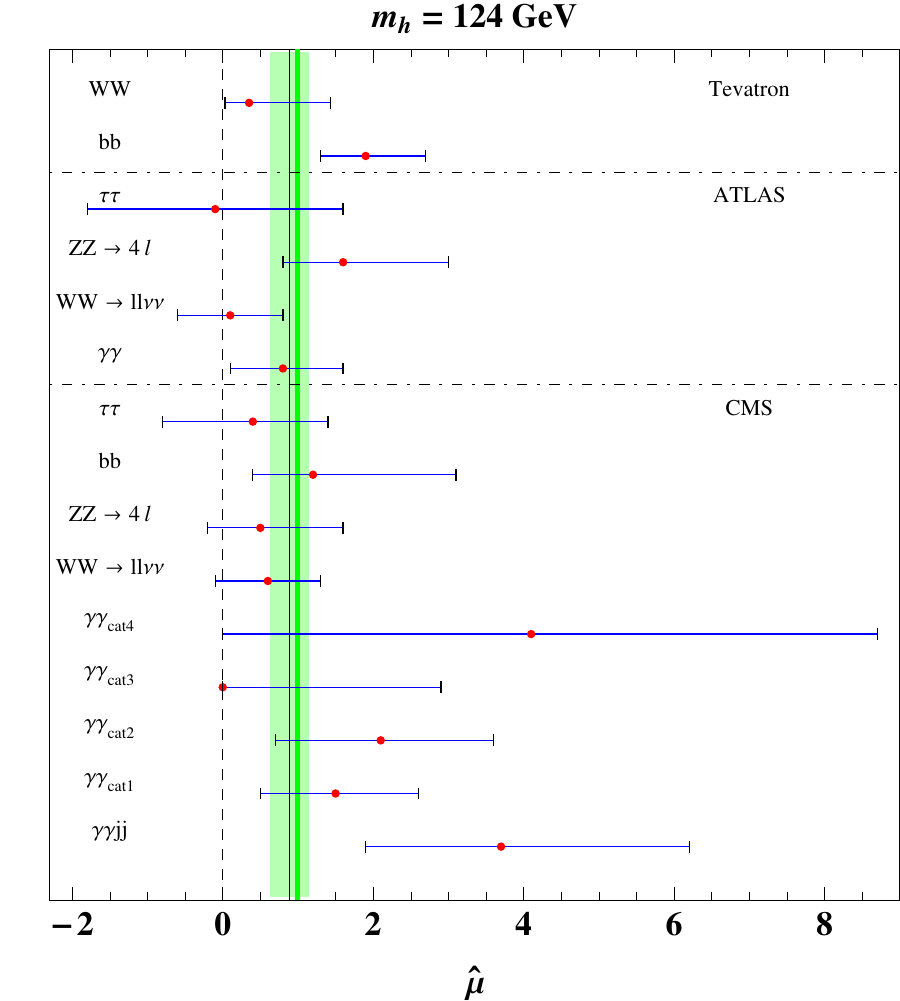}
\includegraphics[width=0.485\textwidth]{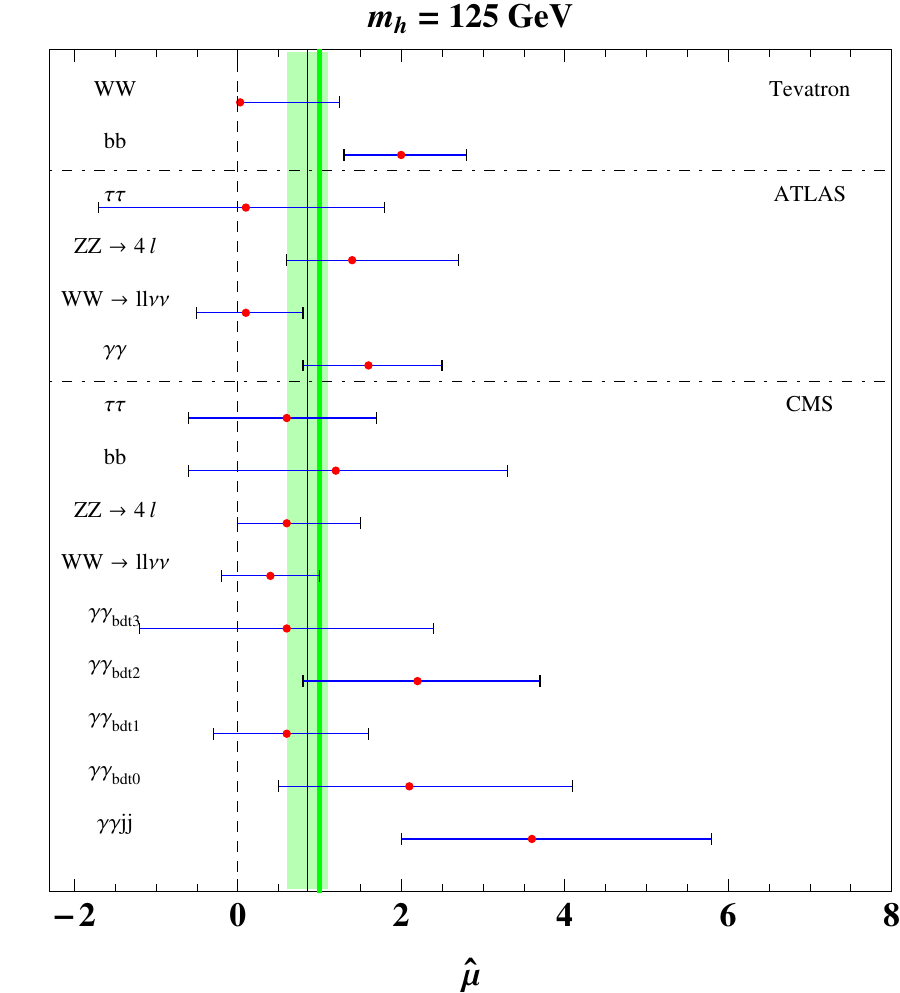}
\caption{\it Pictorial presentation of the data in Table \ref{table:tcorrections} for $m_h=124,125$ GeV. The central value of each signal strength and corresponding error band ($1 \sigma$) is shown.
Also shown is the background only hypothesis as a vertical dashed line, the SM hypothesis as a vertical solid green line at $\hat{\mu} =1$ and a shaded vertical band that corresponds to $\hat{\mu}_c \pm \sigma_c$ as defined in Eq.~\ref{csigma}.
The values of $\hat{\mu}_c$ are $(0.88,0.85)$ for $m_h = (124,125)$ and are shown as a vertical solid black line. These parameters are related to the best fit value of ${\rm Br}_{inv}$ through  ${\rm Br}_{inv}^{\rm min} = 1 - \hat{\mu}_c$.
The results shown here are consistent with the global fit.}
\label{Fig.data}
\end{figure}

For $m_h = 125 \, {\rm GeV}$ we use the public results presented at Moriond 2012 that split the $\gamma \, \gamma$ signal events into
four (multivariate boosted decision tree --BDT) classes that are not identical to the classes used for $m_h = 124 \, {\rm GeV}$. See the relevant experimental papers for the detailed class definition in each case,
but note that the event classes are exclusive (though correlated) and can be combined directly in our $\chi^2$ procedure. Once again correlation coefficients are neglected as they are not supplied, but the effect of 
pseudo-correlations have been examined in Ref. \cite{Espinosa:2012ir} and the fit was found to be stable against randomly chosen correlations. Also we have found in  Section \ref{SMHiggs1} consistent results between two different approaches to the fit of signal-strength parameters: using the individual channels or the combined results. This indicates that neglected correlations do not bias the fit results outside the quoted errors. 

Note that here we use a value of $0.35^{+1.08}_{-0.31}$ for 
the $p\bar{p} \rightarrow W^+ \, W^- \,\, [{\rm CDF \& D 0\! \! \! /}]$ result for $m_h = 124 \, {\rm GeV}$, unlike in Ref. \cite{Espinosa:2012ir}, where we used the same value as for $m_h = 125 \, {\rm GeV}$.  This introduces a small interpolation error,
but allows better agreement with global combined signal strengths reported by the Tevatron collaboration.

\subsection*{Acknowledgments}
We thank A. Djouadi, R. Gon\c{c}alo, A. Juste, M. Martinez, M. Spira, W. Fisher, J. Huston,
V. Sharma and J. Bendavid for helpful communication on related theory and data.
This work has been partly supported by the European Commission under the contract ERC advanced
grant 226371 �MassTeV�, the contract PITN-GA-2009-237920 �UNILHC�, and
the contract MRTN-CT-2006-035863 �ForcesUniverse�, as well as by the
Spanish Consolider Ingenio 2010 Programme CPAN (CSD2007-00042) and the
Spanish Ministry MICNN under contract FPA2010-17747 and
FPA2008-01430. MM is supported by the DFG SFB/TR9 Computational Particle Physics.
Preprints: KA-TP-22-2012, SFB/CPP-12-32, CERN-PH-TH/2012-151.

\end{document}